\documentclass{aa}
\usepackage{graphicx}
\usepackage{txfonts}

\usepackage[dvipsnames,usenames,table]{xcolor}
\usepackage[colorlinks]{hyperref}
\newcommand {\reac}[6] {$\rm\,{}^{#2}\kern-0.8pt{#1}\,({#3}\,,{#4})\,{}^{#6}\kern-0.8pt{#5}\,$}
\newcommand{\HST}{\textit{HST}}

\newcommand{\feh}{\mbox{\rm [{\rm Fe}/{\rm H}]}}

\newcommand{\Msun}{\mbox{$M_{\odot}$}}

\newcommand{\Lsun}{\mbox{$L_{\odot}$}}
\newcommand{\Mi}{\mbox{$M_\mathrm{i}$}}
\newcommand{\Teff}{\mbox{$T_\mathrm{eff}$}}

\newcommand{\lov}{\mbox{$\lambda_\mathrm{ov}$}}
\newcommand{\amlt}{\mbox{$\alpha_\mathrm{MLT}$}}
\newcommand{\Omegac}{\mbox{$\Omega_\mathrm{c}$}}
\newcommand{\omegai}{\mbox{$\omega_\mathrm{i}$}}
\newcommand{\beq}{\begin{equation}}
\newcommand{\eeq}{\end{equation}}
\newcommand{\beqa}{\begin{eqnarray}}
\newcommand{\eeqa}{\end{eqnarray}}

\begin{document} 

  \title{Multiple stellar populations in NGC~1866}

  \subtitle{New clues from Cepheids and colour-magnitude diagram}
    
    \titlerunning{Cepheids and rotation in NGC~1866}

  \author{Guglielmo Costa
          \inst{1}\fnmsep\thanks{\email{gcosta@sissa.it}}
          \and
          L\'eo Girardi \inst{2}
          \and
          Alessandro Bressan \inst{1}
          \and
          Yang Chen\inst{3}
          \and
          Paul~Goudfrooij \inst{4}
          \and
          Paola~Marigo \inst{3}
          \and
          Tha\'ise~S.~Rodrigues \inst{2}
          \and
          Antonio Lanza \inst{1}
          }
          
    \authorrunning{Costa et al.}

  \institute{SISSA, via Bonomea 365, I-34136 Trieste, Italy         
            \and
            Osservatorio Astronomico di Padova -- INAF, Vicolo dell'Osservatorio 5, I-35122 Padova, Italy
            \and
            Dipartimento di Fisica e Astronomia Galileo Galilei, Universit\`a di Padova, Vicolo dell'Osservatorio 3, I-35122 Padova, Italy
            \and
            Space Telescope Science Institute, 3700 San Martin Drive, Baltimore, MD 21218, USA
             }

  \date{Received - ; accepted -}

\hypersetup{
    linkcolor=blue,
    citecolor=blue,
    filecolor=magenta,      
    urlcolor=cyan
}

  \abstract{ 
  We performed a comprehensive study of the stellar populations in the young Large Magellanic Cloud cluster NGC 1866, combining the analysis of its best-studied Cepheids with that of a very accurate colour-magnitude diagram (CMD) obtained from the most recent Hubble Space Telescope photometry.
  We used a Bayesian method based on new PARSEC stellar evolutionary tracks with overshooting and rotation to obtain ages and initial rotation velocities of five well-studied Cepheids of the cluster.
  We find that four of the five Cepheids belong to an initially slowly rotating young population (of $ 176 \pm 5$~Myr), while the fifth is significantly older, either $ 288 \pm 20$~Myr for models with high initial rotational velocity (\omegai~$ \sim 0.9$), or $ 202 \pm 5$~Myr for slowly rotating models.
  The complementary analysis of the CMD rules out the latter solution while strongly supporting the presence of two distinct populations of $\sim$176~Myr and  $\sim$288~Myr, respectively.
  Moreover, the observed multiple main sequences and the turn-offs indicate that the younger population is mainly made of slowly rotating stars, as is the case of the four younger Cepheids, while the older population is made mainly of initially fast rotating stars, as is the case of the fifth Cepheid. Our study reinforces the notion that some young clusters like NGC~1866 harbour multiple populations. This work also hints that the first population, i.e. the older, may inherit the angular momentum from the parent cloud while stars of the second population, i.e. the younger, do not.
}


  \maketitle
%
%

\section{Introduction}

NGC~1866 is one of the most populous young clusters in the Magellanic Clouds. Its location in the northern outskirts of the Large Magellanic Cloud (LMC) disc, little affected by interstellar dust and field stars, makes it one of the most interesting clusters for testing intermediate-age stellar populations and, in particular, stellar evolutionary models with masses close to its turn-off mass of $4-5$~\Msun. This  young cluster contains not only a large number of evolved stars, but also an extremely high number of Cepheid variables \citep[more than 20; ][]{Musella2016}. Early studies of its colour–magnitude diagram (CMD) \citep{Brocato1994, Testa1999, Barmina2002} were concentrated on discussing the efficiency of convective core overshooting in intermediate-mass stars. More recently, the excellent photometry provided by the \textit{Hubble Space Telescope} (\HST) has revealed new surprises, such as the presence of extended main-sequence turn-offs (eMSTO) and a split main sequence (MS). The interpretation of these features are still under debate in the community, but similar features are also observed in other clusters, such as NGC~1844, NGC~1856, and NGC~1755 \citep{Correnti2015, DAntona2015, Milone2016, Bastian2017}. 

A recent study by \citet{Milone2017} demonstrated that these features are real and can result from an unexpectedly high presence of fast rotators in the cluster. Fast rotators were also detected directly, around the cluster turn-off, by spectroscopic measurements of line broadening \citep{Dupree2017}.

Fast rotation is believed to contribute to the eMSTOs in clusters up to much larger ages until about 2~Gyr \citep{Cordoni2018}. There is however intense debate on whether such populous young and intermediate-age clusters also harbour multiple stellar populations, formed at different ages (spanning more than a few Myr) and/or with different initial chemical compositions \citep{Niederhofer2017, Bastian2018, Martocchia2018}.  

In the present work, we aim to reinterprete the available data for NGC~1866 in the light of new PARSEC \citep{Costa2019} stellar models including rotation. Instead of looking at individual features in the CMD, such as the split-MS, eMSTO, and evolved stars, we seek to reinterpret the entire available data; in particular we exploit Cepheids that have accurate pulsational mass determinations. The paper is organized as follows: In Section~\ref{sec:Data_and_methods} we present the NGC 1866 data set and the Cepheid sample. In this section we also summarize the statistical methods used in this study. In Section~\ref{sec:Parsec} we give a brief description of the new physics adopted in the PARSEC code and we present the new evolutionary tracks and isochrone grids. In Section~\ref{sec:Results}, we present the results obtained from the Cepheids analysis, and compare the observed CMD with selected isochrones. Finally, in section~\ref{sec:Disc_and_conc} we discuss the results and draw our conclusions.
%
%
\section{Data and methods}
\label{sec:Data_and_methods}

    \subsection{NGC 1866 photometry}
    \label{sec:NGC1866_data}
        \begin{figure}
            \includegraphics[width=0.45\textwidth]{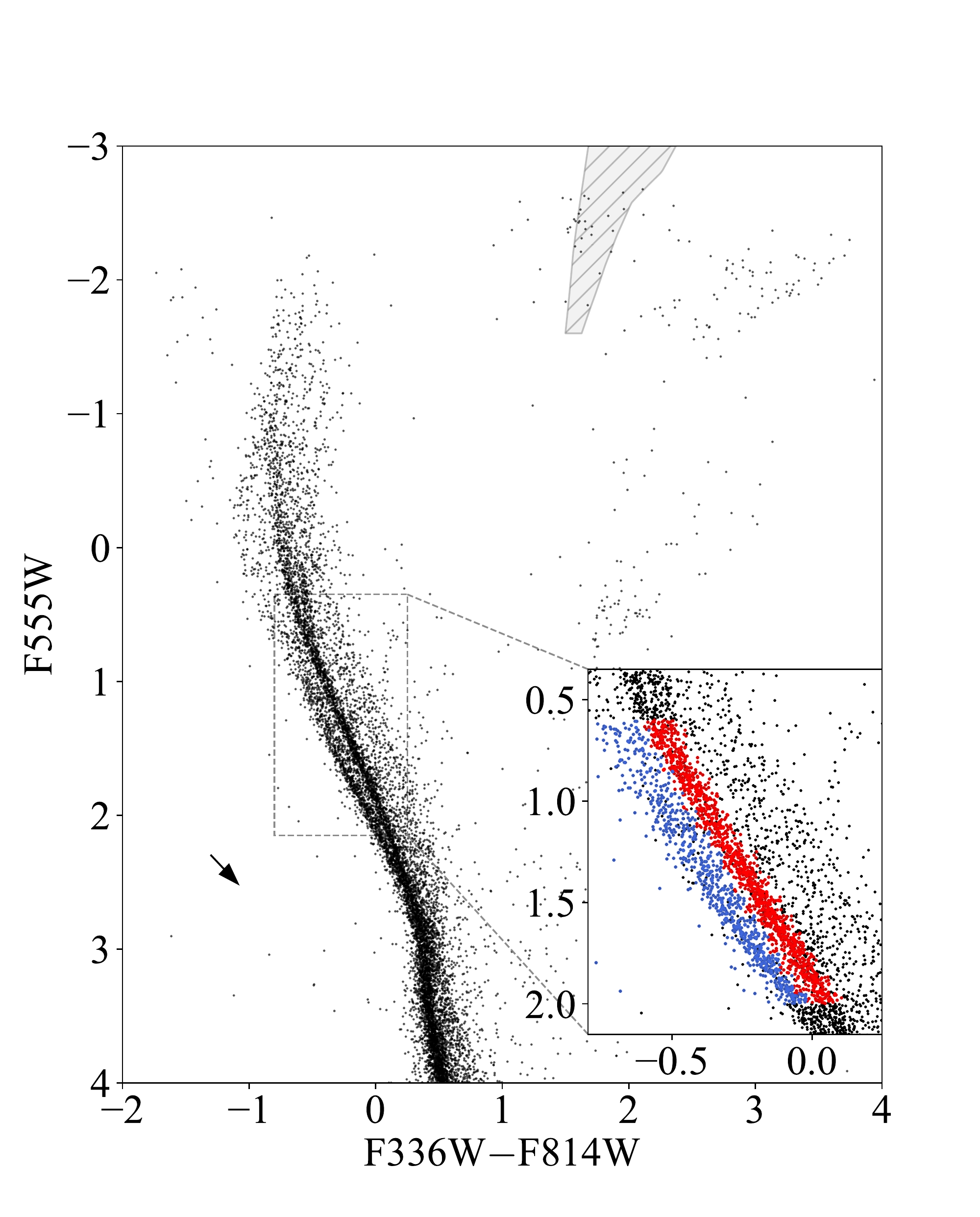}
            \caption{
                F555W vs. F336W $-$ F814 CMD for all the selected stars of the cluster (see text), corrected for the effective magnitude zero points for each band, a true distance modulus of $(m - M)_0 = 18.43$ mag, and a foreground extinction of $A_V = 0.28$ mag \citep{Goudfrooij2018}. The black arrow indicates a reddening vector corresponding to $A_V = 0.1$ mag for reference. The grey shaded region indicates the Cepheids instability strip. The luminosity and  effective temperature of the strip are taken from \citet{Marconi2004} and are coloured with the new YBC bolometric database available at 
                \protect\url{http://stev.oapd.inaf.it/YBC/} \citep{Chen2019}.
                The diagram clearly shows the MS split and the extended MS turn-off. The inset shows a zoom of the MS in which the two MSs are selected and coloured to indicate the bMS (in blue) and the rMS \citep[in red;][]{Milone2017}.
                    }
            \label{fig:cmd}
        \end{figure}
        The NGC 1866 photometric data were acquired through the Ultraviolet and Visual Channel of the  Wide Field Camera 3 (UVIS/WFC3) and the Advanced Camera for Surveys of the Wide Field Camera (ACS/WFC) on board \HST. The cluster was observed in the four pass-band filters F336W, F438W (from WFC3), F555W, and F814W (from ACS). The data were taken in programmes GO-14204 (PI: A. P. Milone) and GO-14069 (PI: N. Bastian), downloaded from the \HST\ archive, and reduced as described in \citet{Goudfrooij2018}. The catalogue comprises $\sim 2\times10^4$ objects. Fig.~\ref{fig:cmd} shows the F555W versus F336W $-$ F814W colour-magnitude diagram (CMD). As suggested by \citet{Girardi2019}, the use of a broad wavelength baseline as colour stretches out the MS and allows us to better identify its features, which in this case are the split-MS and the eMSTO. The inset shows a zoom of the MS region with a subdivision of selected stars with photometric errors smaller than 0.05 mag in all filters, for the blue sequence (bMS) and the red sequence (rMS) shown in blue and red points, respectively. 
        
        As already mentioned, the split-MS is a real feature of the cluster and seems to be present in many young to intermediate-age star clusters. The case of NGC 1866 has been studied recently by many authors \citep[e.g.][]{Milone2017, Goudfrooij2018}, and the common findings are that both rotation and different ages are required to explain the double MS and the eMSTO at the same time. Rotation or the spread in age alone are not able to reproduce the data consistently. Different authors found slightly different values for the metal content in the cluster. \citet{Lemasle2017} in their spectroscopic analysis of a sample of Cepheids in the cluster, found a very homogeneous chemical composition of \feh~=~$-$0.36$\pm0.03$ dex, in agreement with previous measurement of the red giant branch stars by \citet{Mucciarelli2011}.
        The MS presents a third main feature: its extension as a broad strip towards redder colours and brighter magnitudes (the dark points in the inset of Fig.~\ref{fig:cmd}). This feature corresponds to the well-known sequence of nearly equal mass binaries, which in practice spreads a fraction of the stars from the bMS and rMS, up to a maximum upward excursion of 0.75 mag  \citep[see][]{haffner1937}.
    \subsection{Cepheids data}
    \label{sec:Cep_and_Bays}
        \begin{table*}
        \caption{Structural parameters of the Cepheids sample from \citet{Marconi2013}. Averaged \feh\ data from \citet{Lemasle2017}.} 
        \centering
        \begin{tabular}{ccccc}
        \hline\hline 
         Name    & Mass [\Msun]    &  $\log$ L [\Lsun]  &  \Teff [K]    & \feh \\
        \hline
        HV 12197 & 4.6 $\pm$ 0.2   & 3.045 $\pm$ 0.012  & 5950 $\pm$ 12 & $-$0.36 $\pm$ 0.03\\
        HV 12198 & 4.2 $\pm$ 0.1   & 3.10  $\pm$ 0.01   & 6050 $\pm$ 12 & $-$0.36 $\pm$ 0.03\\
        HV 12199 & 3.5 $\pm$ 0.1   & 2.91  $\pm$ 0.01   & 6125 $\pm$ 12 & $-$0.36 $\pm$ 0.03\\
        We 2     & 4.31 $\pm$ 0.15 & 3.00  $\pm$ 0.01   & 5925 $\pm$ 12 & $-$0.36 $\pm$ 0.03\\
        V6       & 4.0 $\pm$ 0.1   & 3.03  $\pm$ 0.01   & 6300 $\pm$ 12 & $-$0.36 $\pm$ 0.03\\
        \hline
        \end{tabular}
        \label{tab:stars_data}
        \end{table*}
        In this work, we adopted two complementary approaches to study NCG 1866. Firstly, we analysed a sample of five Cepheids of the cluster, selected and studied by \citet{Marconi2013}, with a Bayesian statistical method. Later, we fitted the cluster features (the MS split and the eMSTO) with new PARSEC isochrones described in Sec.~\ref{sec:Tracks&Isos} and \ref{sec:GravDark_CMD}. 
        
        The use of Cepheids is convenient for several reasons. First, these stars are characterized by pulsational instabilities that have been extensively studied in recent decades and the physical mechanisms behind their periodical nature are well known. Thanks to very accurate pulsational models it is possible to derive their intrinsic stellar properties, in particular, the pulsational mass, luminosity, photospheric radius, and the \Teff, with a very good precision. 
        Another reason concerns the rotation properties. Since evolved single stars are in general slow rotators, their surfaces are not significantly distorted by the centrifugal forces. Hence, the difference in temperature between the poles and the equator is negligible, and their position in the Hertzsprung-Russell (HR) diagram does not depend on the inclination of the rotation axis with respect to the line of sight. As shown by \citet{Girardi2019} in their Figure 1, non-negligible effects of the rotation in the stellar geometry start to arise for $\omega > 0.5$, where $\omega=\Omega/$\Omegac\ and \Omegac\ is the  angular velocity of rotational break-up.

        Table~\ref{tab:stars_data} lists the intrinsic properties given by the \citet{Marconi2013} analysis, performed by a data fitting procedure with non-linear convective pulsation models. Their models are computed taking into account a mild core overshooting. Their data are multi-filter photometric light curves and radial velocity measurements for the selected Cepheids. The table is complemented with the mean metal content derived by \citet{Lemasle2017} for several Cepheids of NGC 1866.
        
        The data in Table~\ref{tab:stars_data} show that four Cepheids have a
        mass slightly greater than 4~\Msun, and one has a mass of $M=3.5$~\Msun. Assuming that all the stars belong to the same stellar population, it is surprising to see such a large range of masses for evolved  stars, taking into account the associated small errors in the pulsational mass determination. One possible solution is that HV12199, the star with lower mass, lost part of its mass during its previous evolution \citep{Marconi2013}. However this is difficult to explain in the framework of a common mass loss rate formulation for the red giant stars. Alternatively, Cepheid HV12199 could be the result of a particular binary evolution history. However 
        this would require a fine tuning for HV12199 to loose only a small fraction (about 0.5~\Msun)  of its external envelope that, for an evolved star of 4~\Msun, is of about 3~\Msun. 
        Another possibility is that HV12199 really belongs to a population older than the other Cepheids and hence it has a lower post MS mass. We investigate this possibility in more detail in the following sections.
    \subsection{Bayesian statistical analysis}
    \label{sec:Bay_stat}
        We analysed the Cepheids data by means of new stellar evolutionary models (described in the following section) using a full Bayesian statistical analysis. We used the PARAM \citep{dasilva2006, Rodrigues2014, Rodrigues2017} code to obtain the posterior joint probability density functions (JPDFs) that depend only on three parameters, which are the metallicity \feh, age, and initial rotation rate (\omegai) of the stars. As prior functions we adopted the following: (1) a flat prior on age, that is, all ages between minimum and maximum values of $3.0\times10^7$~yr and $9.8\times10^{8}$~yr are assumed to be equally likely; (2) similarly, a flat prior on the initial angular rotation rate parameter \omegai, between values of 0 and 0.95; and (3) a mass distribution given by the initial mass function (IMF) from \citet{Kroupa2002}.
      
        In Section~\ref{sec:Results} we present the results of the analysis for each star, and we discuss the possible belonging to the same stellar population of the Cepheids, hence assuming a common age and metallicity content. In that case, we combine the different JPDFs to obtain the corrected JPDF (cJPDF). The cJPDF of each star shares the age and metal content with the other without having any constraints on their initial rotation rates, \omegai. \citet{Costa2019} have provided a detailed description of the statistical method and computation of the cJPDFs.
        
%
%
\section{PARSEC models with rotation}
\label{sec:Parsec}

    \subsection{New prescriptions}
    \label{sec:implemented_physics}
        \begin{figure}
            \includegraphics[width=0.48\textwidth]{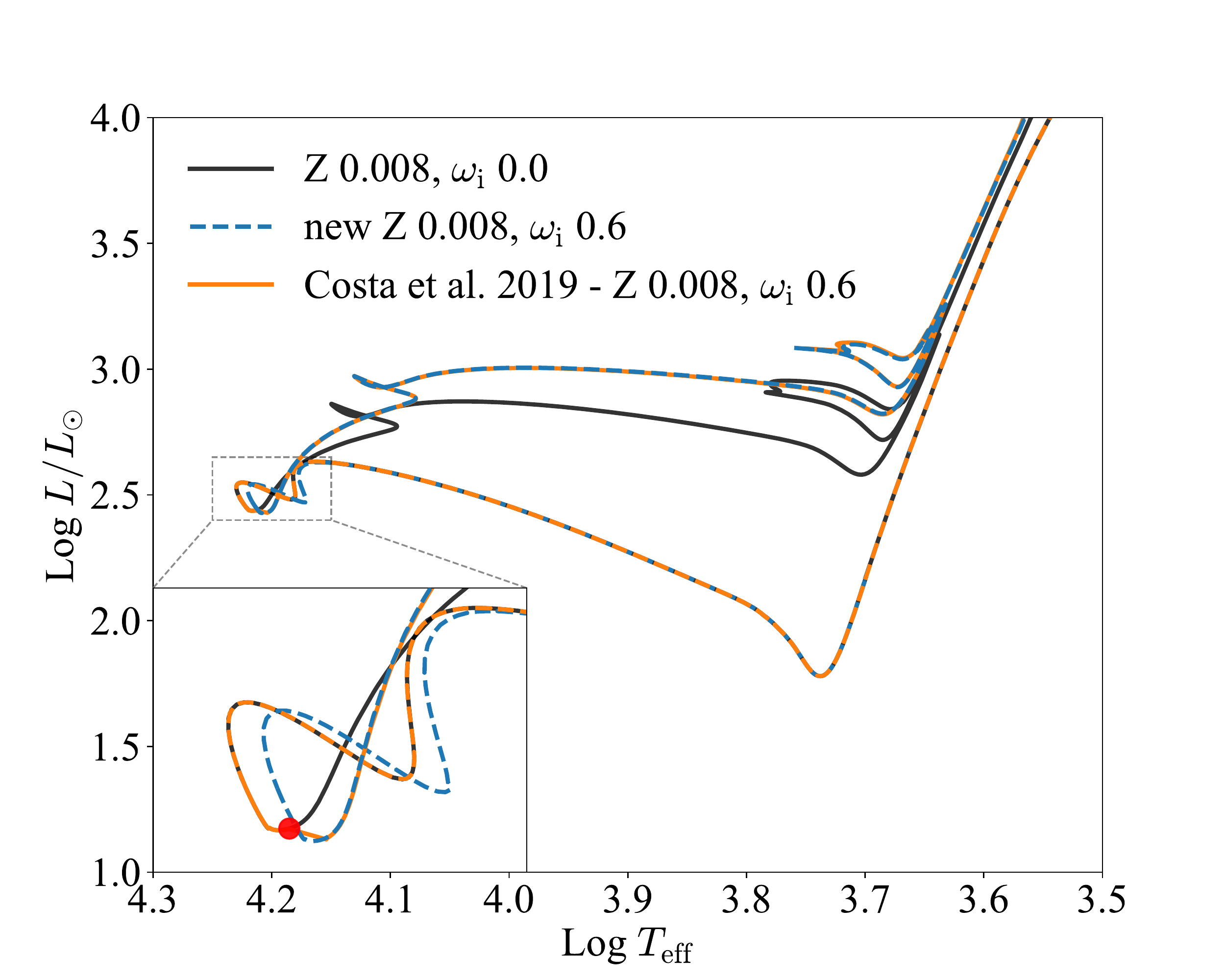}
            \caption{Comparison between tracks of a 4~\Msun\ star. The black line is the non-rotating model, while the dashed blue and the solid orange lines are the new and old rotating tracks, respectively. A zoom of the ZAMS region is shown in the inset to emphasize the differences between the old and new methods of introduction of the angular velocity (see text).
            }
            \label{fig:old_new_tracks}
        \end{figure}
        \begin{figure*}
            \includegraphics[width=0.48\textwidth]{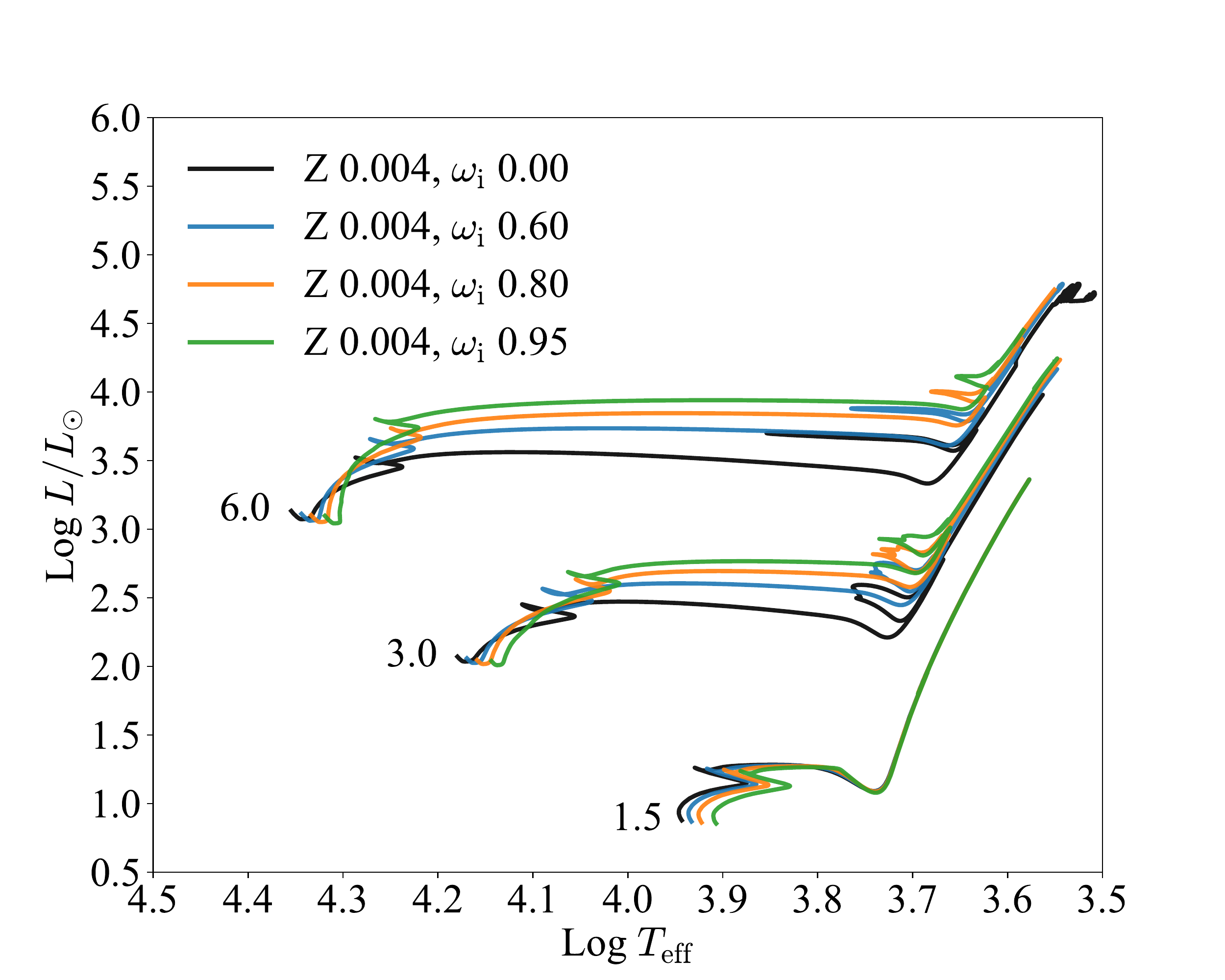}
            \includegraphics[width=0.48\textwidth]{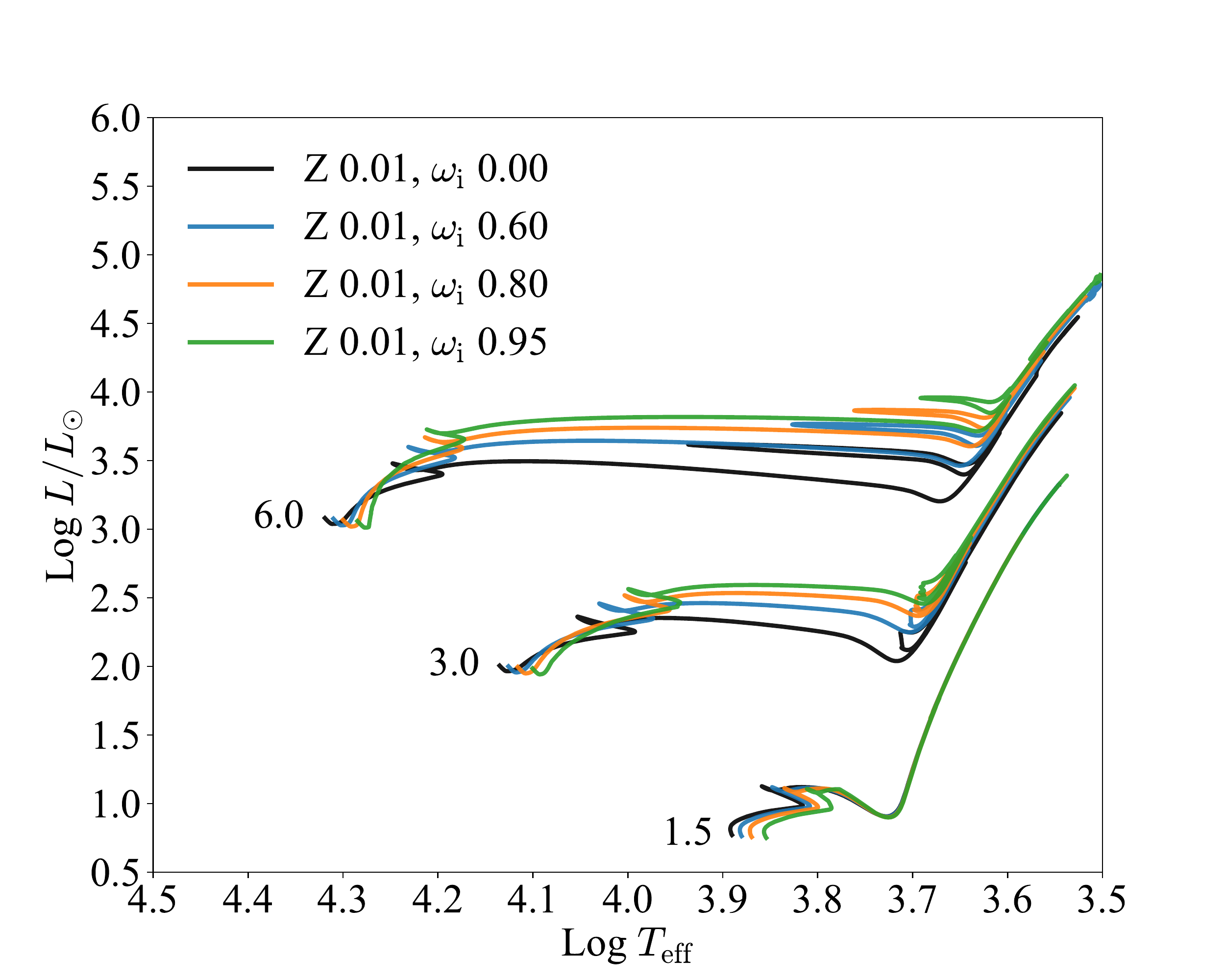}
            \caption{Selected evolutionary tracks from our sets. In different colours different rotation rates (\omegai~= 0.0, 0.60, 0.80, 0.90), for three different masses (1.5, 3, and 6 \Msun). The left panel shows tracks with a selected metallicity of Z~=~0.004, while the right panel shows tracks with Z~=~0.01.
                    }
            \label{fig:selected_tracks}
        \end{figure*}
        \begin{figure}
            \includegraphics[width=0.48\textwidth]{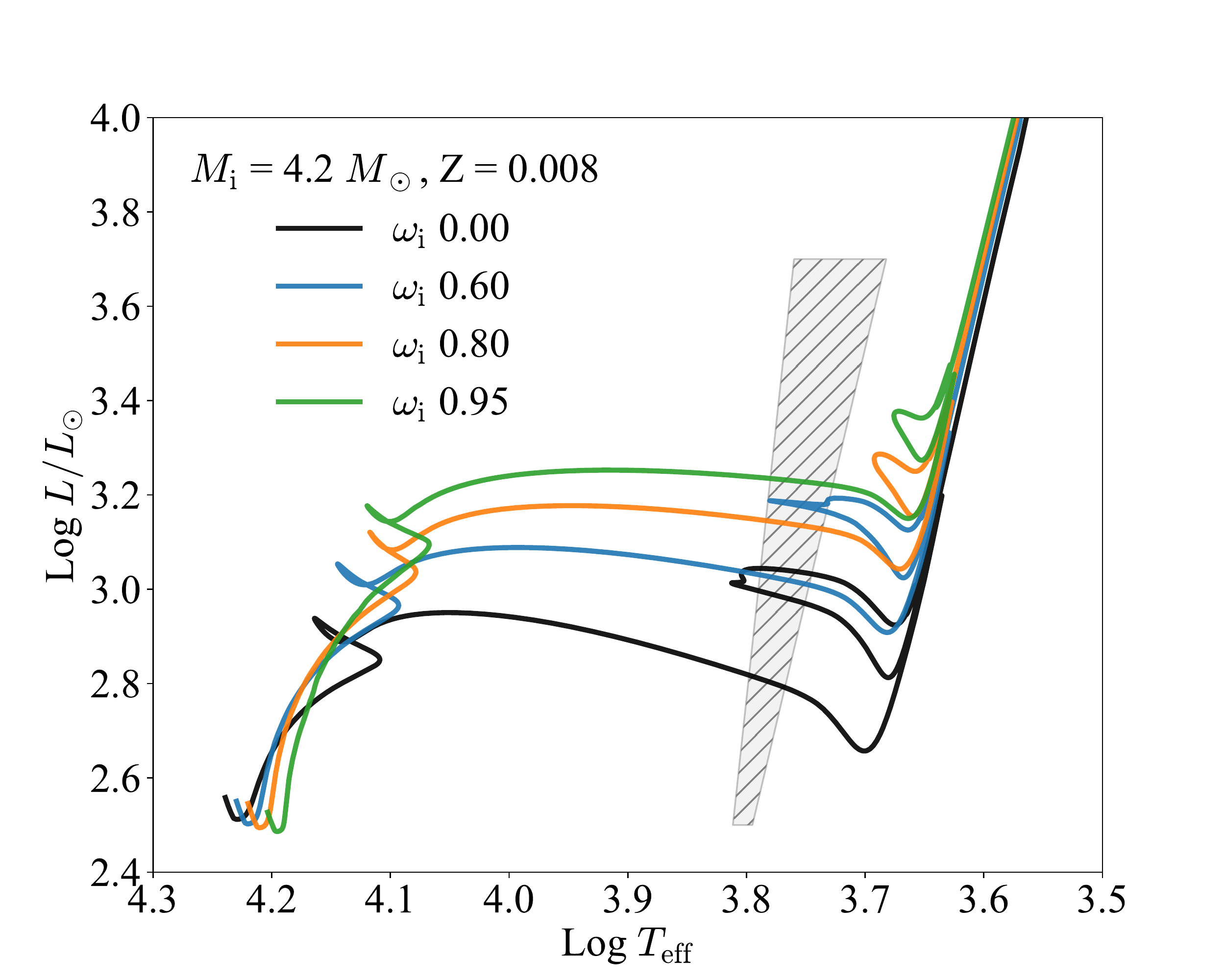}
            \caption{Evolutionary tracks of a 4.2~\Msun\ star with different initial rotation rates (\omegai) in different colours. The grey shaded area indicates the Cepheids instability strip \citep{Marconi2004}.
                    }
            \label{fig:tracks_cepheid}
        \end{figure}
        We used the PARSEC V2.0 code \citep{Costa2019} to compute models of rotating stars. The other features of the code are extensively described in \citet{Bressan2012, Bressan2015}, \citet{Tang2014},  \citet{Chen2014, Chen2015} and \citet{Fu2018}. In the present work we implemented new physics that is briefly described in the following.
        
        We updated the nuclear reaction network, which now consists of 33 isotopic elements from hydrogen to zinc, including the reverse reactions of the $\alpha$-captures. In total there are 72 reactions.
The second main update concerns mass loss. In previous releases of PARSEC evolutionary tracks (without rotation), mass loss was activated for stars more massive than 12~\Msun, which are the only stars to be significantly affected by this process during their main nuclear burning phases. However, in the case of rotating stars, mass loss must also be taken into account at lower masses for the following reasons. First of all, there is an enhancement of mass loss due to the lower effective gravity, which is reduced by the centrifugal forces. Secondly, mass loss removes angular momentum from the star, assuring the stability of the angular momentum transport and the evolution of the star. This is particularly true in cases of fast rotating stars (e.g. $\omega \geq 0.90$). To include this effect in stellar models we used the prescription provided by \citet{Heger2000a}, who modify the mass loss rate as follows:
        \begin{equation}
            \dot{M}(\omega) = \dot{M}(\omega = 0) 
            \left(\frac{1}{1-v/v_\mathrm{crit}}\right)^\xi, \quad \mathrm{ with }\quad \xi = 0.43
            \label{eq:mass.loss}
        ,\end{equation}
        where $\dot{M}(\omega = 0)$ is the mass loss rate in case of no rotation, computed using the prescriptions provided by \citet{deJager1988} for low- to intermediate-mass stars, and with the prescriptions described in \citet{Chen2015} for massive stars. The quantity $v$~is the surface tangential velocity of the star, and $v_\mathrm{crit}$ is the break-up velocity, that is 
        \begin{equation}
            v^2_\mathrm{crit} = \frac{Gm}{r} (1-\Gamma_\mathrm{E})  \,\,\,\, ,
            \label{eq:crit.vel}
        \end{equation}
        where $\Gamma_\mathrm{E}$ is the Eddington factor.
        During the evolution, stars with high initial rotation rates (\omegai~$ \geq 0.90$) may reach the critical rotation (typically at the end of the MS). In that case, the surface effective gravity at the equator is zero owing to the centrifugal forces, and the most external layers become detached from the star. This is usually called mechanical mass loss. As suggested by \citet{Georgy2013}, we may expect that this phenomenon to happen mainly in the equatorial region of the star, and the supercritical layers escape in such a way as to maintain the surface at the critical velocity or slightly below it. The mass loss by winds (also called radiative mass loss) computed so far using Eq.~\ref{eq:mass.loss}, may not be enough to extract the momentum required for keeping the external shells below the critical velocity. To treat this, the code computes the mechanical mass loss starting from the angular momentum excess found in the supercritical shells. This excess is the difference between the actual angular momentum of the supercritical shells and their critical angular momentum, defined as $L_\mathrm{c} = \Omega_\mathrm{c}\,I$, where $I$ is the momentum of inertia of each shell. From the excess momentum we estimated the mass that should be removed to keep the star below its critical rotation.
        Because of numerical difficulties, we define a maximum angular rotation rate, that is $\omega_\mathrm{max} = 0.998$. After these calculations the code selects the largest mass loss between the two and removes it from the star.
        We carefully treated the mass loss enhancement and the mechanical mass loss implementation in the code, taking particular attention to the angular momentum conservation over time. At each time step, the sum of the current angular momentum of the star plus the total momentum lost by the wind is equal to the initial angular momentum given to the star. In this work, we do not take into account for longitudinal anisotropy of the wind and coupling with the magnetic field of the star. Other prescriptions for the mass loss enhancement by rotation are provided by \citet{Maeder2000} and \citet{Georgy2011, Georgy2013}.
       
	    With respect to the previous version of the code \citep{Costa2019}, we adopted a new strategy to assign the initial angular rotation rate to the star. In the former method the initial rotation rate is assigned when the star reaches the zero age main sequence (ZAMS) in a single time-step. With the new method the angular velocity is increased in time, starting from few time-steps before the ZAMS (about 40 time-steps, corresponding to $\sim 2$~Myr in case of a 4~\Msun\ star with \omegai~$=$~0.60). If the desired \omegai\ is reached before the ZAMS, we did not allow $\omega$ to grow above that value. After the ZAMS is reached this condition is relaxed, and the angular velocity is let free to evolve. This new approach allowed 
        us to have tracks that smoothly reach the new `rotating' ZAMS without big jumps between the non-rotating pre-main sequence (PMS) and the rotating MS. This is shown in Fig.~\ref{fig:old_new_tracks} for a 4~\Msun\ star. With the new method a 4~\Msun\ star at the `rotating' ZAMS with \omegai~=~0.60 has ingested an angular momentum that differs less than the 0.3 per cent with respect to the previous method. The following evolution is not affected by this different approach.
        \begin{figure*}
            \includegraphics[width=0.48\textwidth]{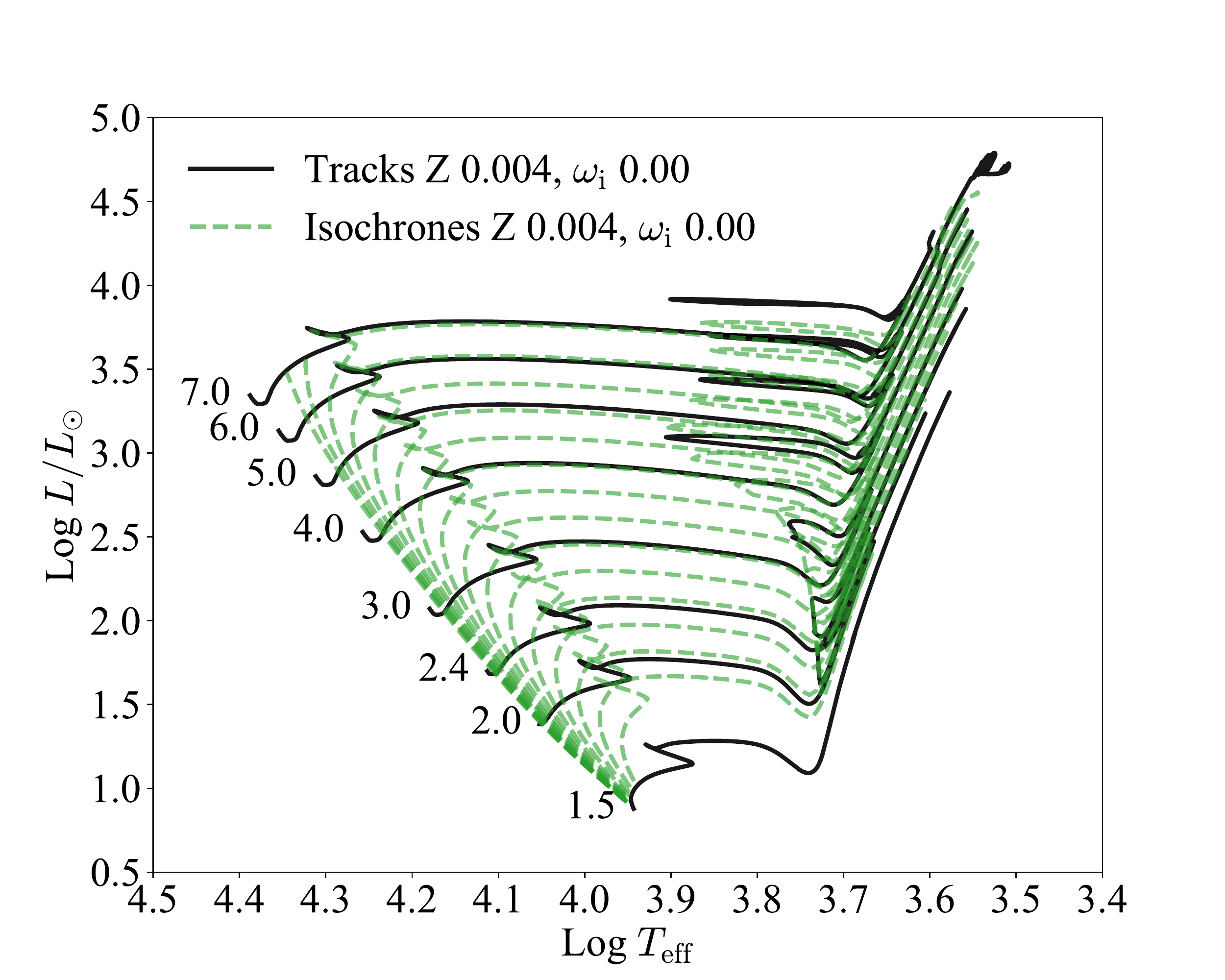}
            \includegraphics[width=0.48\textwidth]{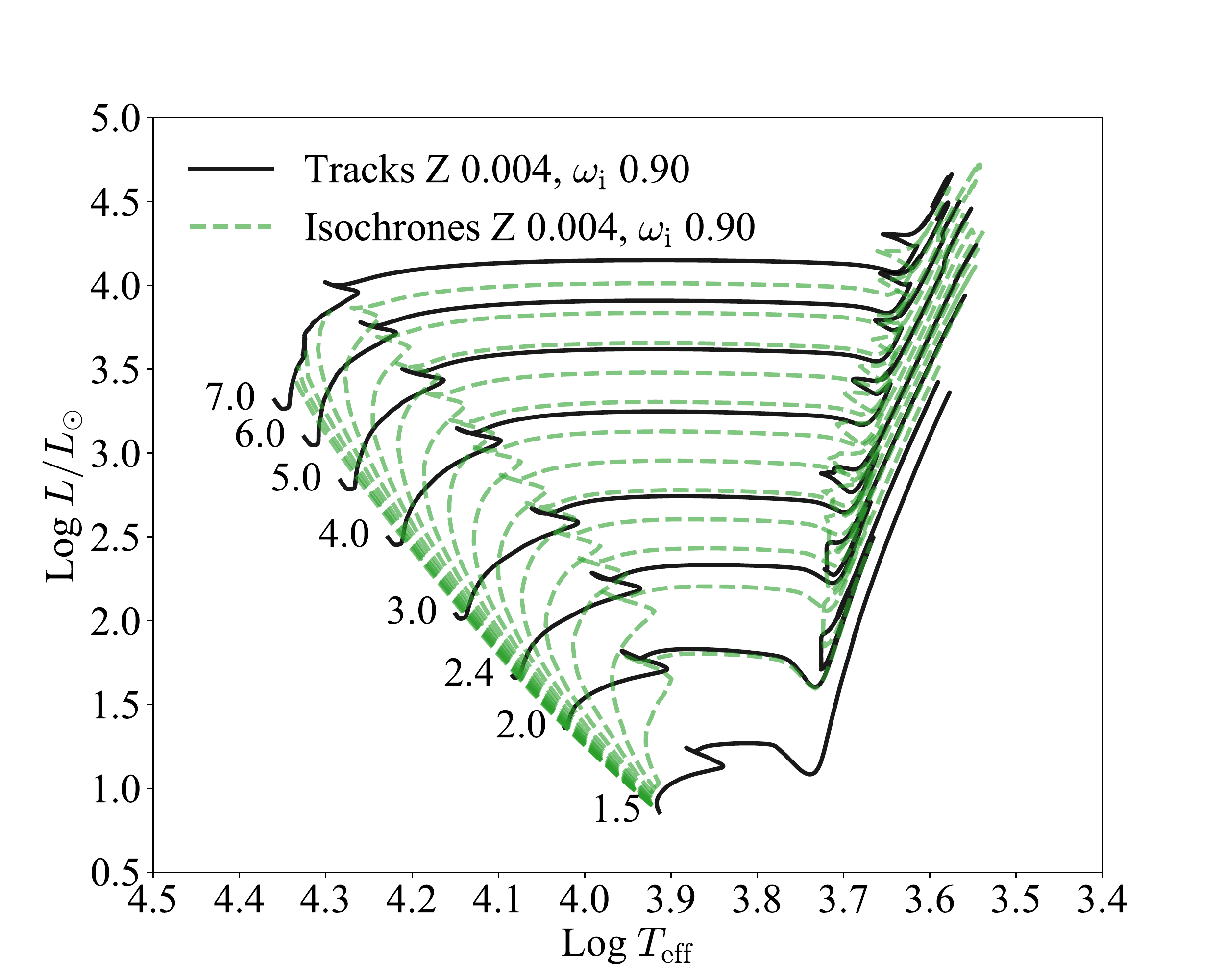}
            \newline
            \includegraphics[width=0.48\textwidth]{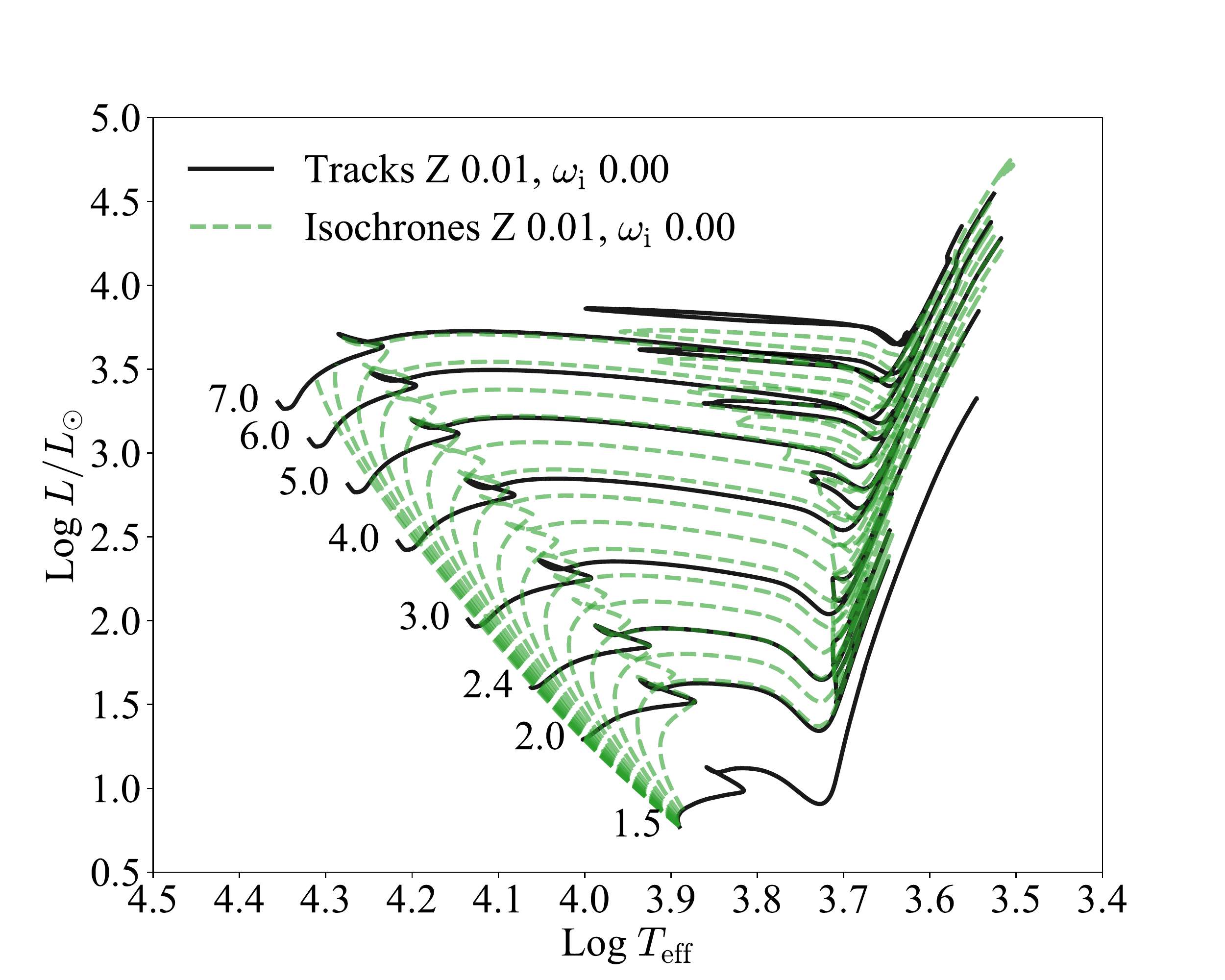}
            \includegraphics[width=0.48\textwidth]{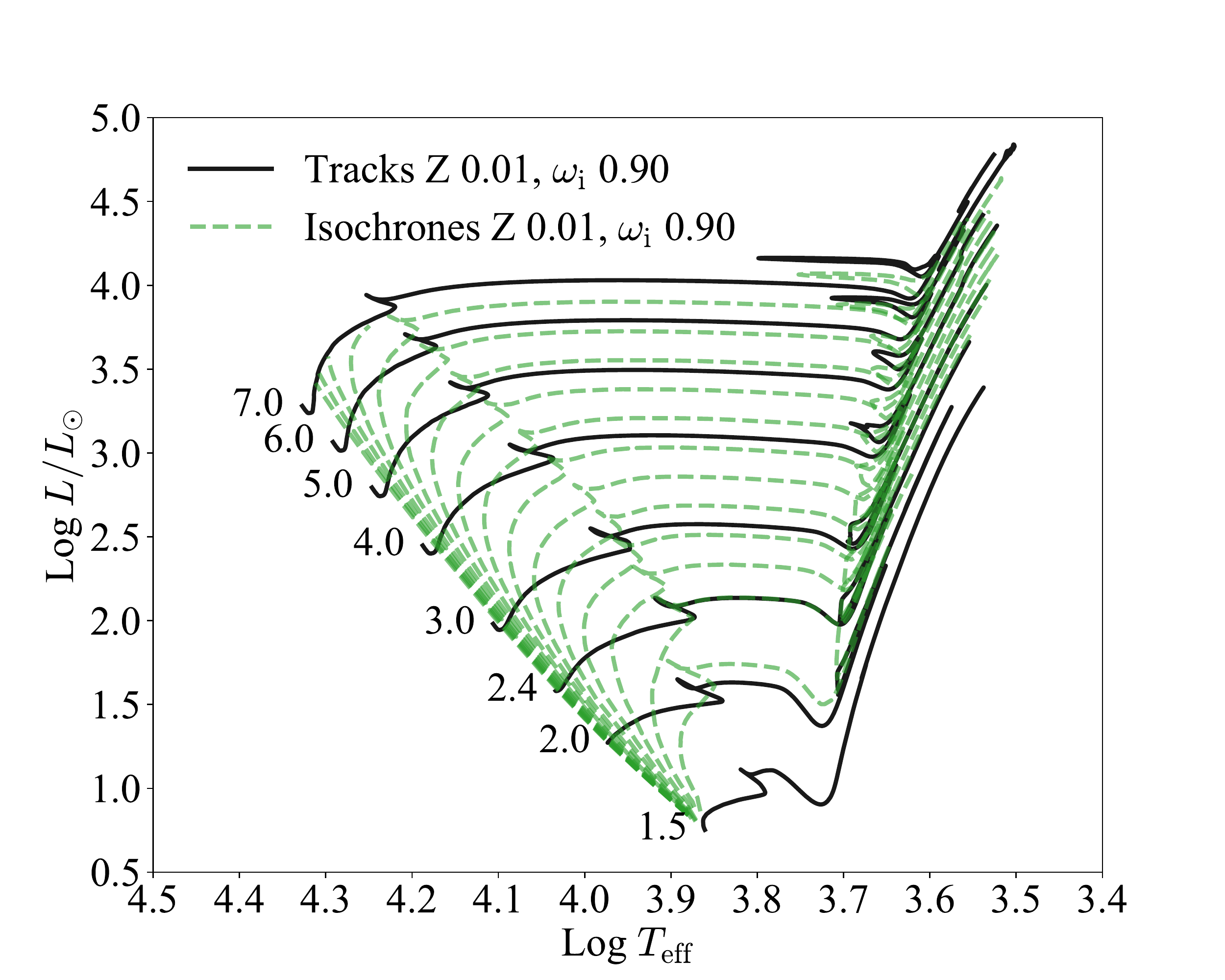}
            \caption{
            Selected evolutionary tracks (black solid lines) over-plotted with the corresponding isochrones (green dashed lines) used in this work. The left-hand column panels show tracks with \omegai~=~0.0, while the right-hand column panels show those with \omegai~=~0.90. The top row panels show tracks with Z~=~0.004 and the bottom panel shows those for Z~=~0.01. The tracks cover the mass range from 1.5 to 7~\Msun. The isochrones are equally spaced in $\log t$ and cover the age range from 30 to 980~Myr.
                    }
            \label{fig:tracks_isocs}
        \end{figure*}
    \subsection{Evolutionary tracks and isochrones}
    \label{sec:Tracks&Isos}

        We computed the new sets of evolutionary tracks with a mass range  spanning from 1.5~\Msun\ to 7~\Msun. We used solar-scaled mixtures based on \citet{Caffau2011} solar composition, where values of initial metal content are Z~=~0.004, 0.006, 0.008, 0.01 and the respective values of helium content are Y~=~0.256, 0.259, 0.263, 0.267. These values and the adopted fixed mixing length parameter, \amlt~=~1.74, were obtained from the solar calibration performed by \citet{Bressan2012}. The magnetic braking and surface magnetic effects have not been implemented in PARSEC yet.
        
        For each metallicity set, we computed tracks with varying initial rotation rate values (\omegai~=~0, 0.30, 0.60, 0.80, 0.90, 0.95). All the tracks are computed with a fixed core overshooting efficiency parameter, \lov~=~0.4, as suggested by \citet{Costa2019}. The evolutionary were computed with mass loss, implemented as described in Sec.~\ref{sec:implemented_physics}, to handle cases with high rotation rate ($\omega > 0.80$). The mass loss does not influence the MS evolution in case of low mass and slow rotation rates. For instance, in case of a 1.5~\Msun\ star with \omegai~=~0.30, the mass at the base of the red giant phase is 0.01 per cent smaller than its initial mass; we found similar values regardless of the metallicities considered. 
        Fig.~\ref{fig:selected_tracks} shows a comparison between models computed with different initial rotation rates, for models with selected masses and metallicity. Depending on the mass, the effect of rotation is slightly different. In the case of low mass stars (1.5~\Msun), the prevailing effect induced by rotation is geometrical distortion, which depending on the $\omega$, takes the track to run in the HR diagram at lower temperatures (during the MS) with respect to the non-rotating case.
        In the giant phase, the surface angular velocity of the star drops down owing to the conservation of the angular momentum, and the rotating track runs almost superimposed on the non-rotating track.
        In the case of intermediate-mass stars, the mixing induced by rotation starts to play a more and more important role as the mass and the rotation rate increase. 
        The rotational mixing, acting in the radiative regions of the stars, provides fresh fuel to the burning core and transports the processed material to the stellar surface. 
        The extra mixing causes the stars to be more luminous and to build up bigger He cores at the end of the MS than the corresponding non-rotating models. Thus, the tracks of rotating stars run on the Hertzsprung gap at higher luminosities than non-rotating stars, mimicking luminosities of more massive non-rotating stars. 
        Rotating stars live longer in the MS phase and for instance, in the case of a 6~\Msun\ star model (with $Z$~=~0.004), the MS lifetimes are 62, 74, 82, and 89~Myr for \omegai~=~0.0, 0.60, 0.80, and 0.95, respectively. 
        
        The core He-burning (CHeB) phase is also influenced by the rotation, for both the different mass of the helium core (built during the MS phase) and the enhanced mixing acting during the phase. 
        We note that from our models, in the HR diagram the position and extension of the blue loops -- typical features of intermediate-mass stars during the CHeB phase --  are affected by rotation. As \omegai\ increases, the blue loops become
        less extended and the CHeB lifetimes decrease. 
        In particular, for the case of the 6~\Msun\ star model (with Z~=~0.004), the CHeB lifetimes are 6, 5, 4, and 3 Myr for \omegai~=~0.0, 0.60, 0.80, and 0.95, respectively.
        The extension of the blue loop due to different rotations plays a crucial role in determining the number of Cepheids in the CHeB phase. In Figure~\ref{fig:tracks_cepheid} we show selected tracks of a 4.2~\Msun\ star with different rotation rates and with Z~$= 0.008$. 
        The chosen mass is the average mass of the stars in Table~\ref{tab:stars_data} excluding HV12199. The grey shaded area in the plot indicates the Cepheids instability strip, and it can be easily seen how the blue loops of the tracks cross the strip only once or multiple times, depending on the \omegai. From the comparison, we may already expect that the number of Cepheids belonging to initially slow rotating populations, with a turn-off mass and a metallicity similar to those selected for the plot, should be higher than the number of Cepheids that belong to an initially fast rotating population. This argument is discussed further in Sec.~\ref{sec:Disc_and_conc}.
        
        To obtain finer grids of tracks to perform our analysis, we interpolated the sets in metallicity, mass, and \omegai. Details on the method of
interpolation are provided in \citet{Costa2019} and references therein. This method allowed
        us to obtain well-behaved isochrones. Figure \ref{fig:tracks_isocs} shows an example of HR diagrams of four selected evolutionary tracks sets over-plotted with the corresponding isochrones selected in age. These sets of tracks are used as inputs for the Bayesian analysis.

    \subsection{Gravity darkening and colour-magnitude diagram}
    \label{sec:GravDark_CMD}
        \begin{figure}
            \includegraphics[width=0.48\textwidth]{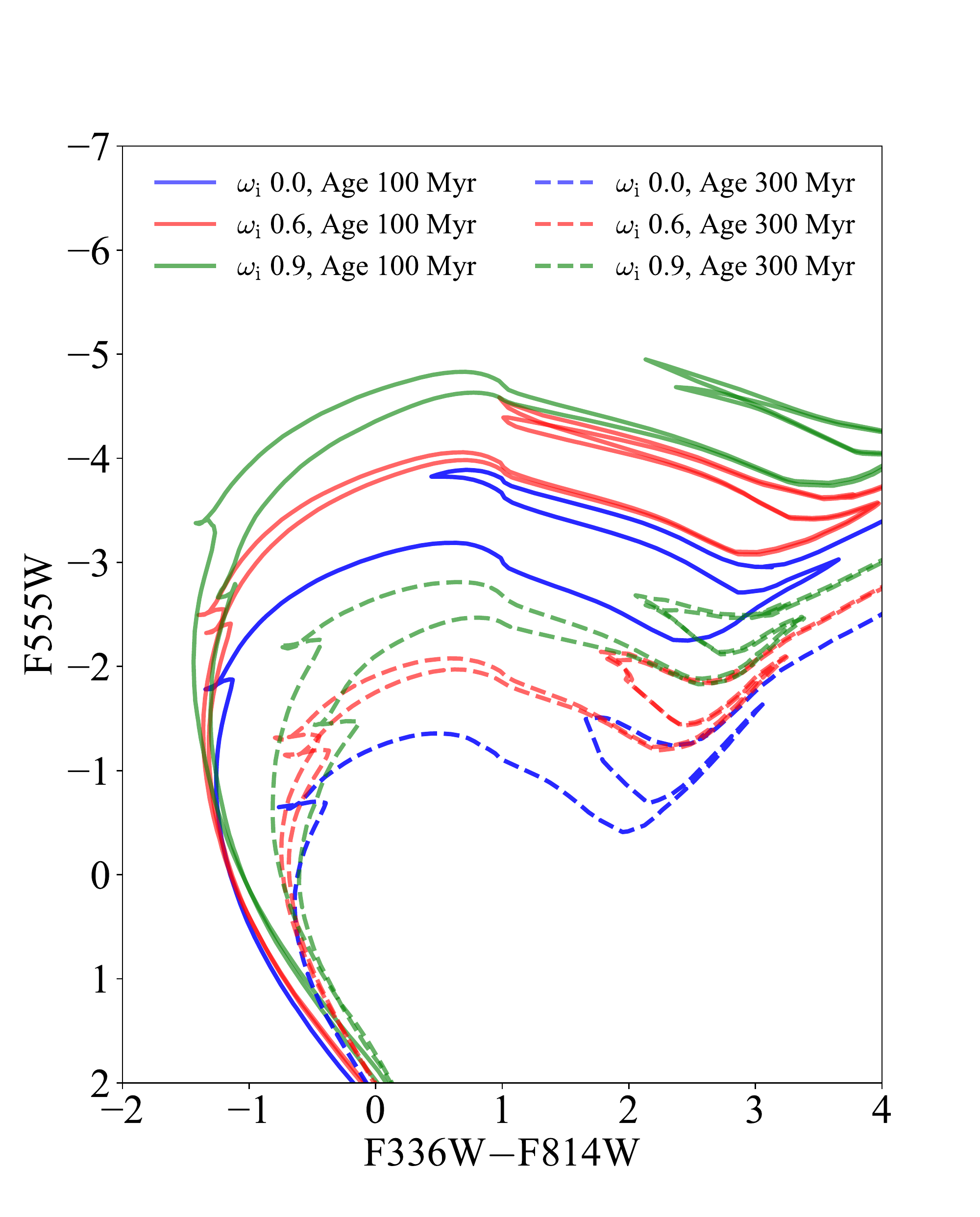}
            \caption{Colour-magnitude diagram of F555W vs. F336W~-~F814W. Effects of gravity darkening on isochrones with different combinations of rotation rates and ages are shown. The solid lines indicate isochrones with the selected age of 100~Myr; the dashed lines indicate those with about 300~Myr. The blue, red, and green lines correspond to \omegai~=~0.00, 0.60, and 0.90, respectively. The metallicity is Z~=~0.006 for all the isochrones. There are two cases for the inclination angles: pole on (\textit{i}~=~0$^\circ$) and edge on (\textit{i}~=~90$^\circ$), which are the brightest and faintest isochrones, respectively.
            }
            \label{fig:isoc_gravity}
        \end{figure}
        As discussed before, rotation induces a distortion of the star shells, which become more and more oblate as the $\omega$ increases. At the critical velocity (\Omegac) the equatorial radius of the star is 1.5 times bigger than the polar velocity. This effect is due to centrifugal forces that reduce the effective gravity along the surface, depending on the co-latitude angle, $\theta$ (with $\theta = 0^{\circ}$ aligned with the rotation axis). Since the local effective temperature is proportional to the local effective gravity, the \Teff$(\theta)$ of a rotating star is not constant along the surface. This effect is known as gravity darkening and was firstly described by \citet{vonZeipel1924}. Such dependence of the \Teff\ on the co-latitude, introduces a new variable in the computation of the total flux emitted by a star, i.e. the inclination angle, $i$, of the star rotation axis with respect to the observer line of sight.
        
        To compute the isochrones in the \HST /WFC3 photometric system, we used the TRILEGAL code \citep{Girardi2005, Marigo2017}, which has recently been updated by \citet{Girardi2019} and \citet{Chen2019} to include the effects of the gravity darkening on rotating stars. The equations of \citet{espinosalara2011} were adopted to calculate the emitted flux, depending on the current surface $\omega$ and inclination angle of the star. The new tables of bolometric corrections (BC) used by TRILEGAL are part of the YBC database\footnote{An on-line version of the tool can be found at \url{http://stev.oapd.inaf.it/YBC/}.} and interpolating routines by \citet{Chen2019} Figure~\ref{fig:isoc_gravity}  shows an example of selected isochrones with different initial angular velocity and with two different ages (100~Myr and 300~Myr). The plot shows the effect of the gravity darkening depending on different rotations. The most evident features are the very different brightness and colour shown by highly rotating stars in the turn-off. If stars in a cluster do not have a preferred angle of inclination, but a distribution of various inclination, high rotating stars could populate such region of the HR diagram creating the eMSTO, which is a feature commonly observed in young to intermediate-age
        star clusters \citep{Milone2017}. The possible alignment of stellar spins in a cluster \citep{Corsaro2017} is still controversial \citep{Mosser2018}.

%
%
\section{Results}
\label{sec:Results}
    
    \subsection{Bayesian analysis}
    \label{sec:res_bayes}
        \begin{figure*}
            \includegraphics[width=0.45\textwidth]{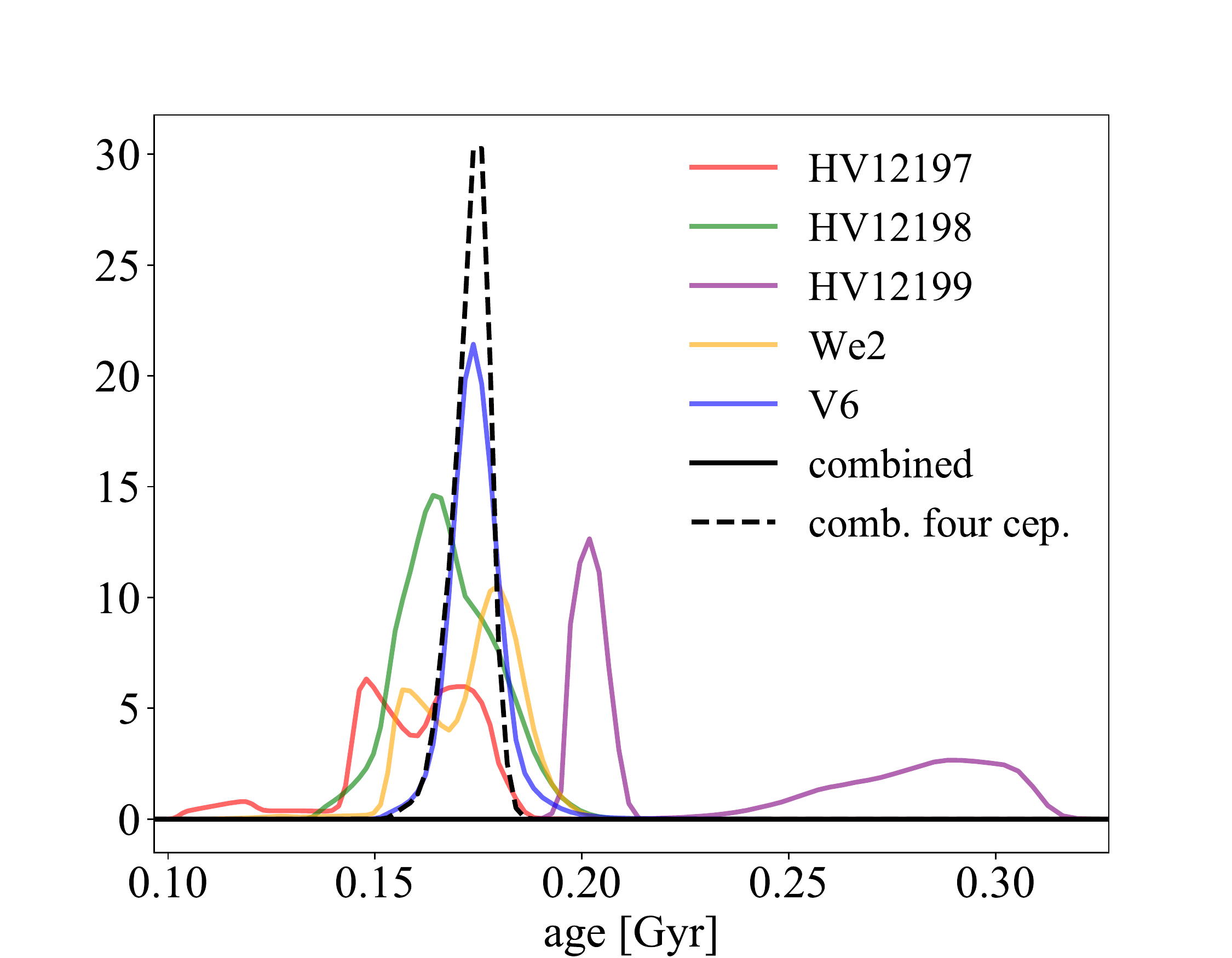}
            \includegraphics[width=0.45\textwidth]{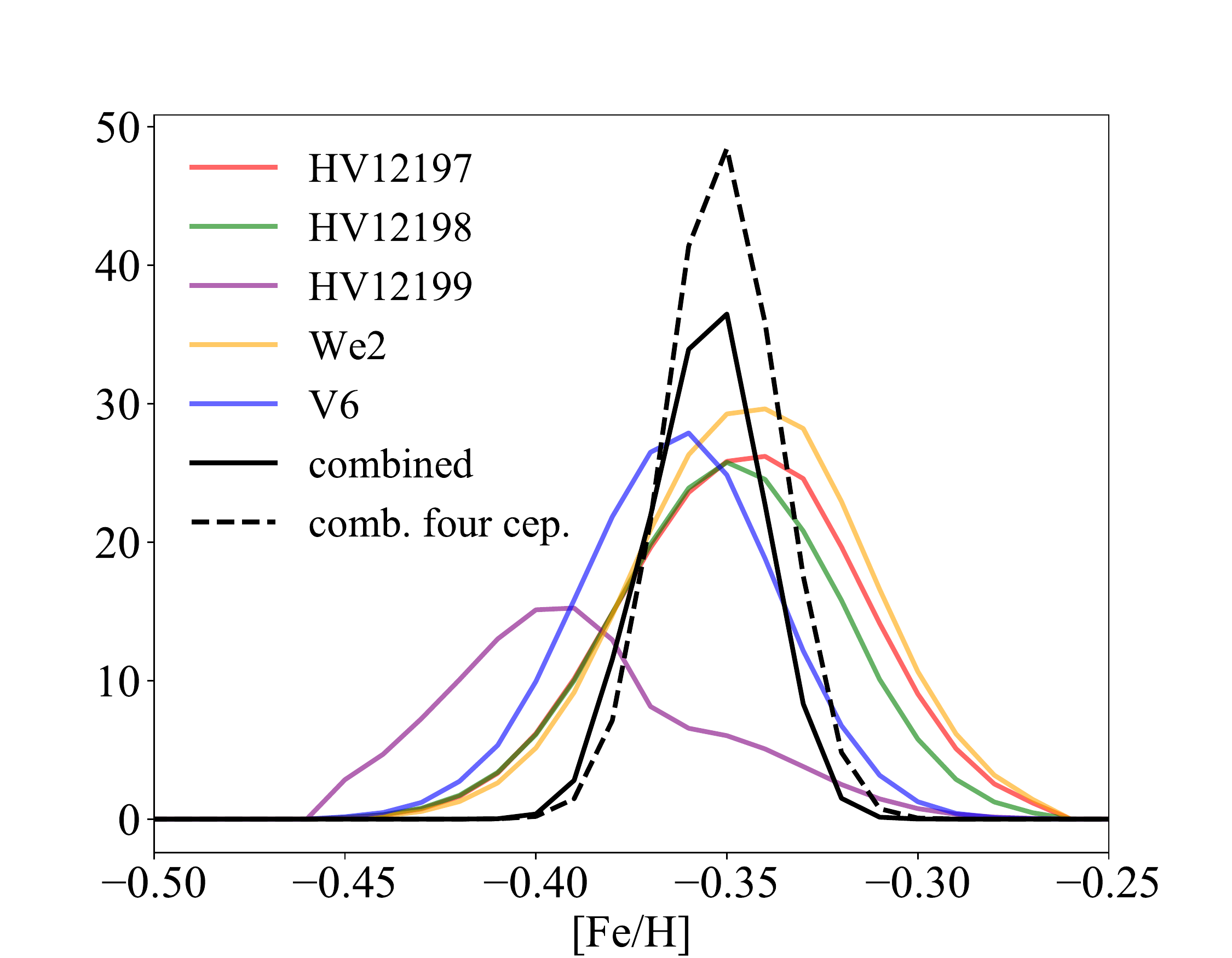}
            \caption{
                Marginalized JPDFs on age (left-hand panel) and on metallicity (right-hand panel) of each Cepheid. The solid black line indicates the combined PDF of all stars; the dashed line indicates the CPDF of four stars (as described in the text).
                    }
            \label{fig:combined_mpdf}
        \end{figure*}
        \begin{figure}
            \includegraphics[width=0.48\textwidth]{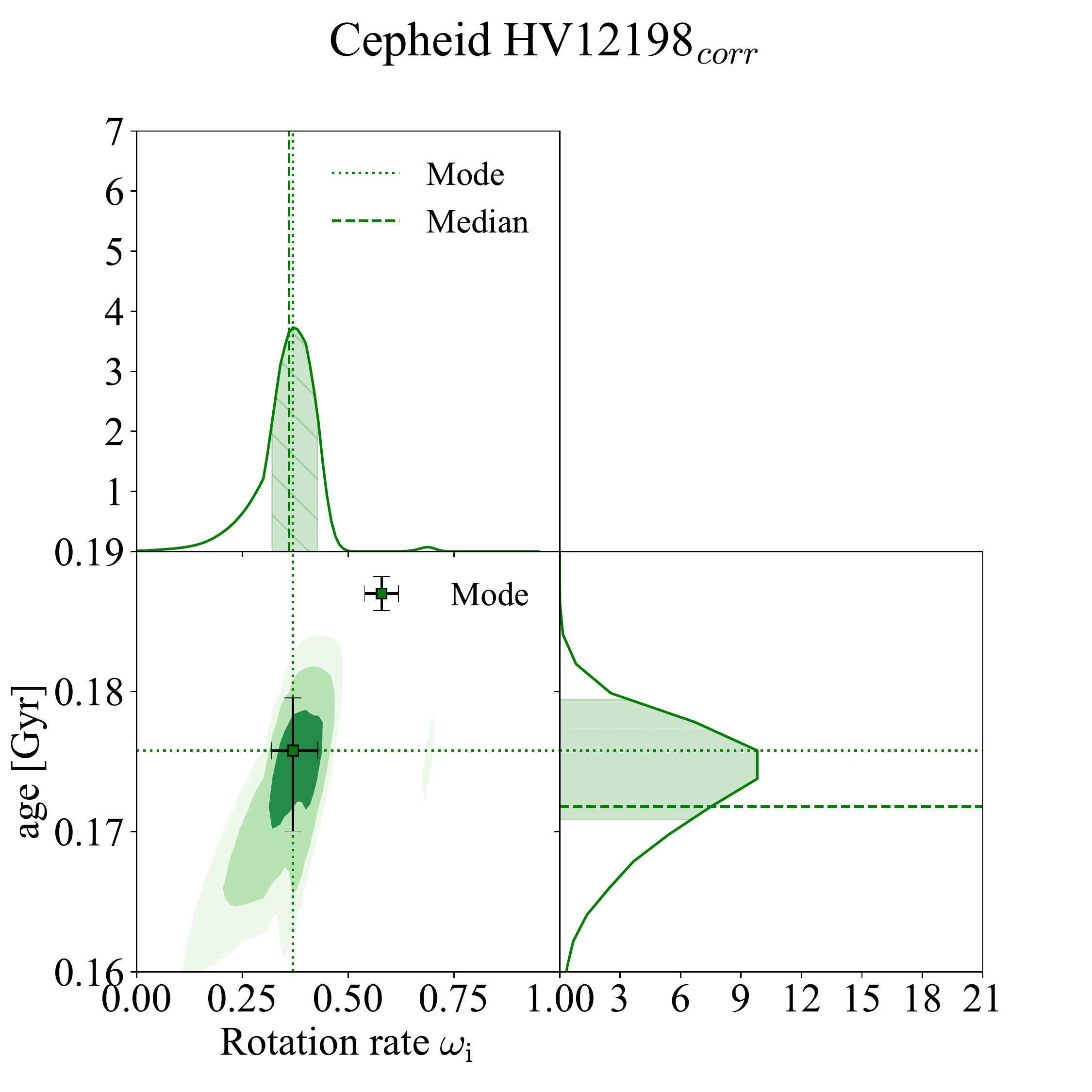}
            \caption{
                Example of selected 2D-cJPDF as a function of age and \omegai\ for the Cepheid HV12198. In the left-hand bottom panel the JPDF is shown; in the left-hand top and right-hand bottom panels the marginalized 1D-PDF of \omegai\ and age, respectively, are shown. The square and error bars in the left-hand bottom panel indicate the mode and the CIs. The coloured contours indicate different levels of the JPDF as indicated in the text. The dashed and dotted lines are the mode and median of the distributions. The shaded areas in the marginalized plots indicate the 68 per cent CI of the two PDFs. 
                    }
            \label{fig:HV19198_corr}
        \end{figure}
        \begin{table} 
            \caption{Resulting values of the initial angular rotation rates and the age for each Cepheid.} 
            \centering 
            \begin{tabular}{lccc} 
                \hline\hline
                Cepheid Name        &       \omegai         & Age [Myr] &  \feh\\ 
                \hline
                HV12197& 0.00$_{-0.00}^{+0.24}$ & 148$_{-3}^{+26}$ & -0.34$_{-0.04}^{+0.02}$ \\
                HV12197$_{corr}$& 0.31$_{-0.16}^{+0.05}$ & 176$_{-6}^{+4}$ & -0.35$_{-0.02}^{+0.01}$ \\
                HV12198& 0.37$_{-0.20}^{+0.07}$ & 164$_{-9}^{+14}$ & -0.35$_{-0.03}^{+0.03}$ \\
                HV12198$_{corr}$& 0.37$_{-0.05}^{+0.06}$ & 176$_{-6}^{+4}$ & -0.35$_{-0.02}^{+0.01}$ \\
                We2& 0.00$_{-0.00}^{+0.24}$ & 180$_{-19}^{+5}$ & -0.34$_{-0.04}^{+0.02}$ \\
                We2$_{corr}$& 0.20$_{-0.12}^{+0.09}$ & 176$_{-6}^{+4}$ & -0.35$_{-0.02}^{+0.01}$ \\
                V6& 0.20$_{-0.10}^{+0.10}$ & 174$_{-6}^{+6}$ & -0.36$_{-0.03}^{+0.02}$ \\
                V6$_{corr}$& 0.18$_{-0.08}^{+0.11}$ & 176$_{-6}^{+4}$ & -0.35$_{-0.02}^{+0.01}$ \\
                \hline 
                HV12199$_\mathrm{CS-1}$& 0.13$_{-0.09}^{+0.03}$ & 202$_{-5}^{+3}$ & -0.40$_{-0.03}^{+0.01}$ \\
                HV12199$_\mathrm{CS-2}$& 0.89$_{-0.06}^{+0.06}$ & 288$_{-23}^{+17}$ & -0.36$_{-0.04}^{+0.02}$ \\
                \hline
            \end{tabular} 
            \label{tab:cep_results} 
        \end{table}
        \begin{figure*}
            \includegraphics[width=0.48\textwidth]{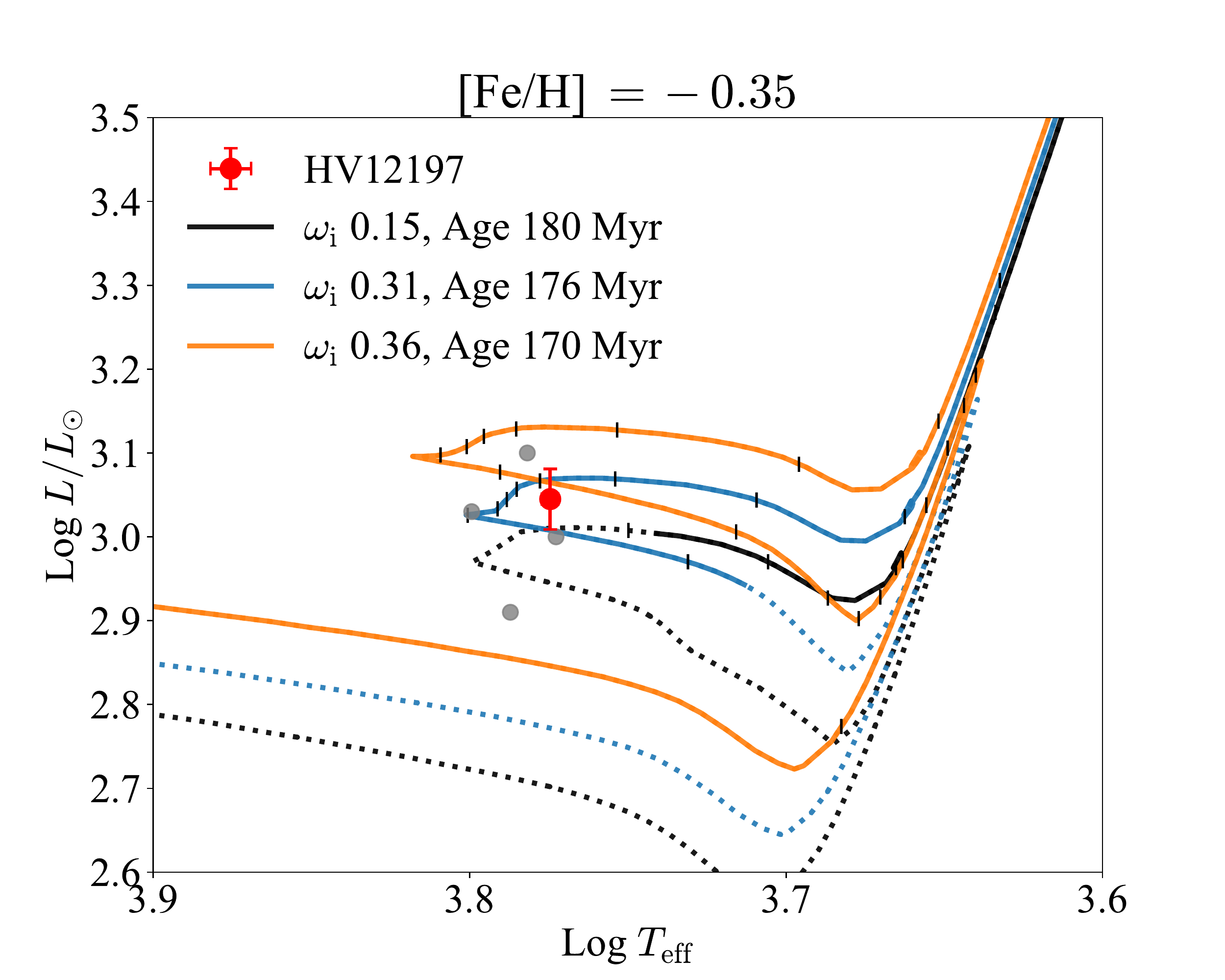}
            \includegraphics[width=0.48\textwidth]{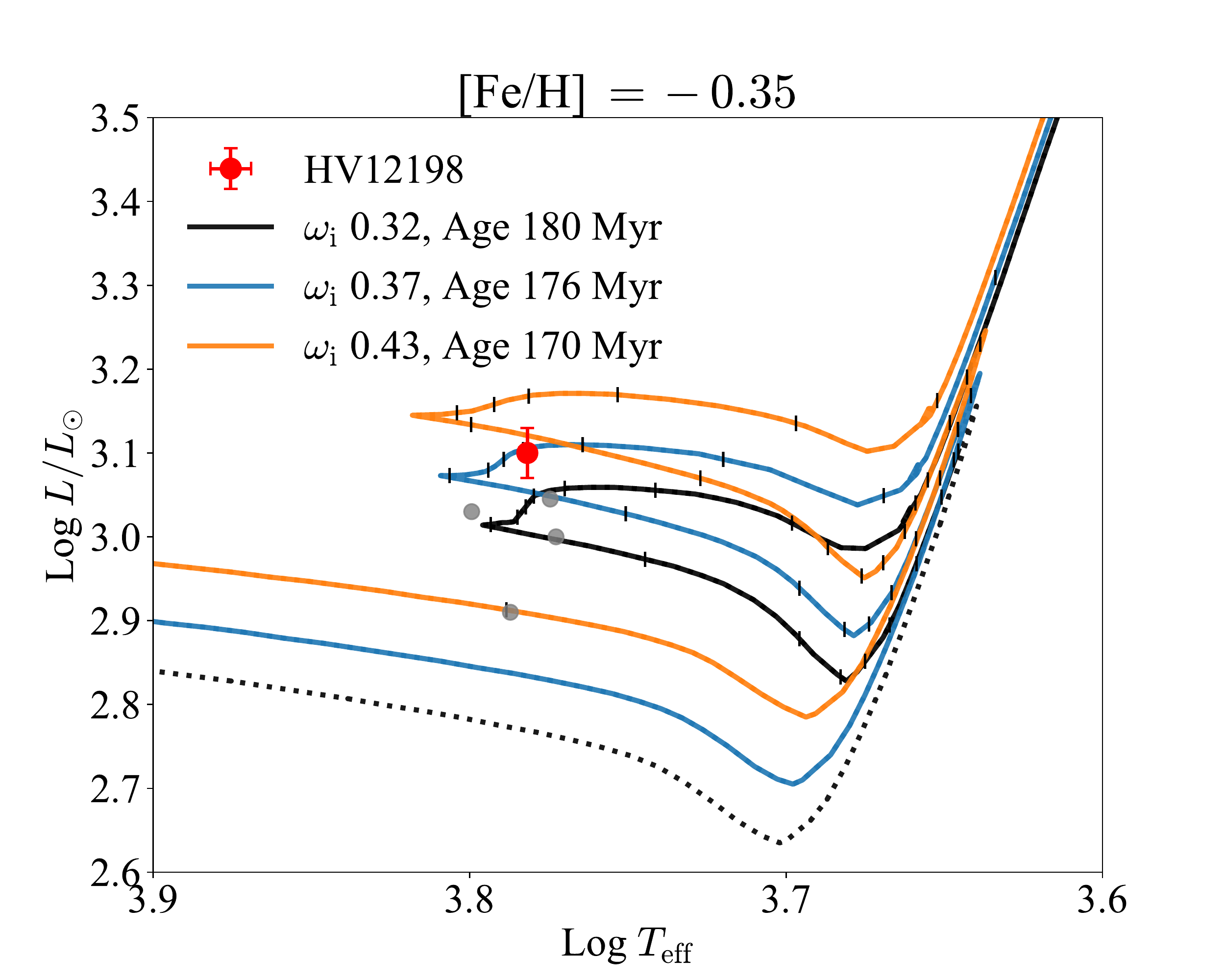} \\
            \includegraphics[width=0.48\textwidth]{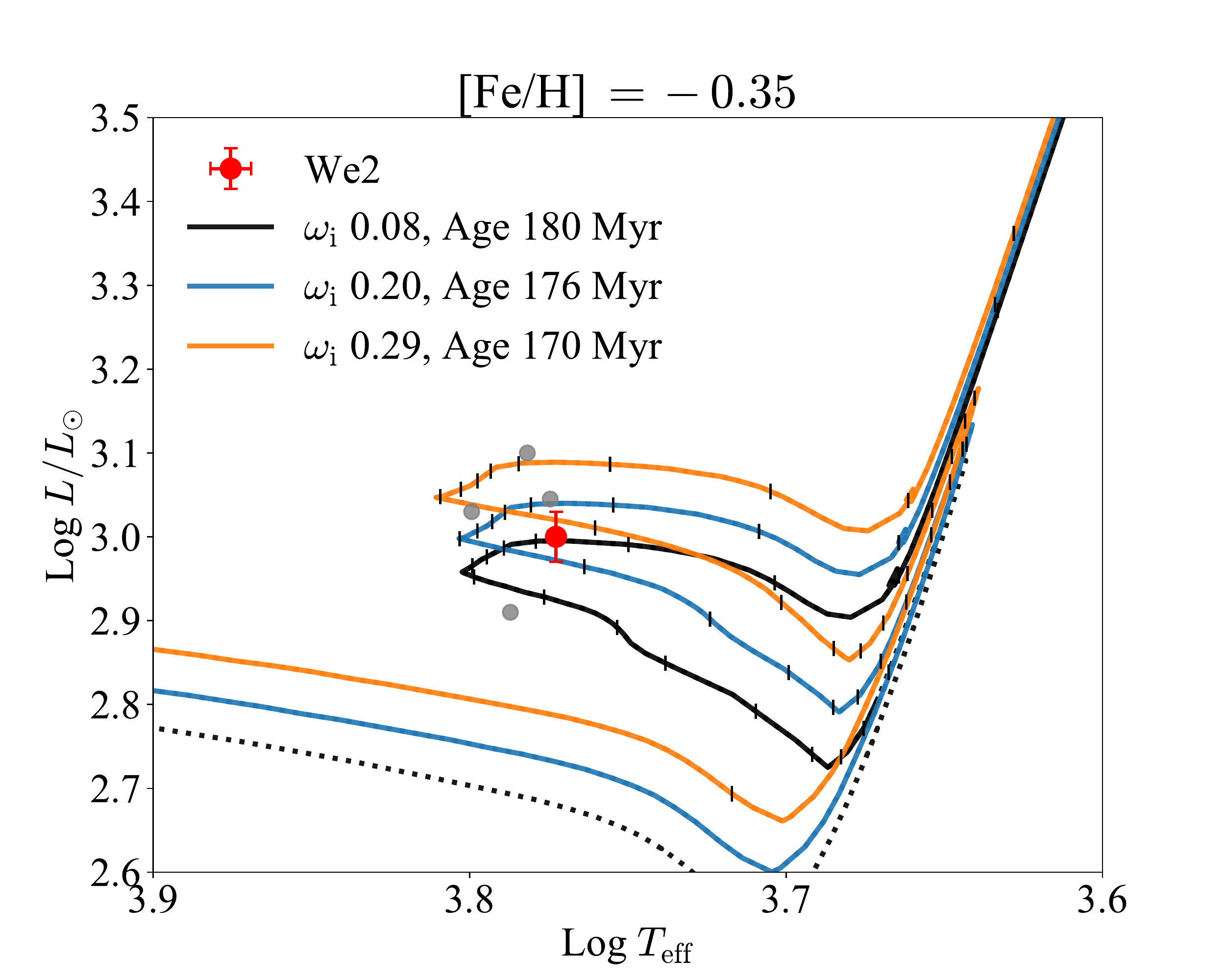}
            \includegraphics[width=0.48\textwidth]{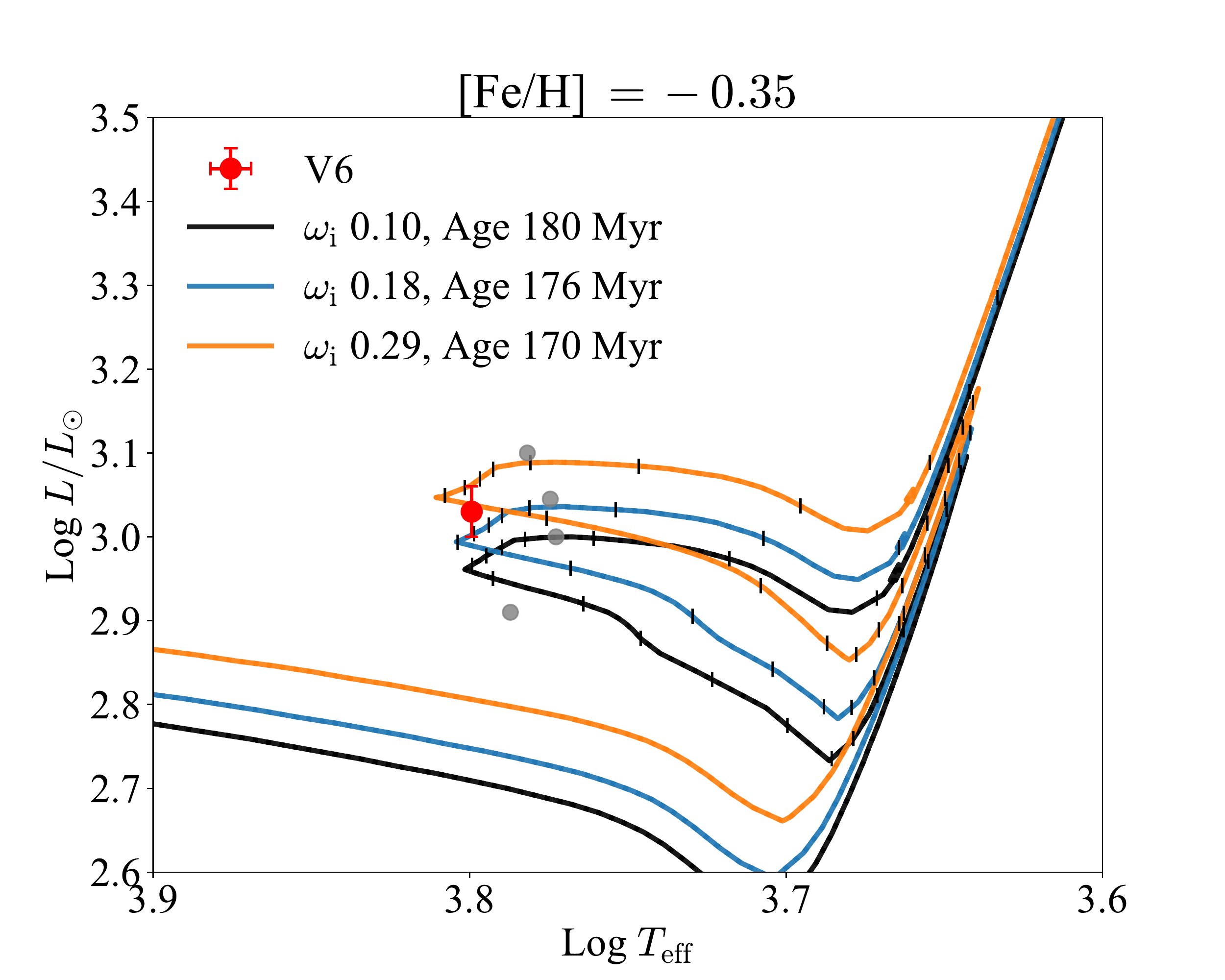}
            \caption{
                Comparison between the Cepheids data and isochrones selected from the best values obtained from the Bayesian analysis. In each panel the red point with the error bars indicate the selected Cepheid; the grey points  indicate the other stars. The error bars are plotted using 3$\sigma$. The blue isochrone represents the best-fitted values from the analysis; age and \omegai\ are indicated in Table~\ref{tab:cep_results}. The black and orange lines are the most and less luminous isochrones within the CIs, respectively. The continuous lines indicate the mass within 3$\sigma$ of the errors in Table~\ref{tab:stars_data}; the dotted lines indicate the part of the isochrones with a mass that is outside the 3$\sigma$ interval.
                To highlight the different evolutionary speed along isochrones, we plot black vertical markers at intervals of $\Delta M$~=~0.01~\Msun.
                    }
            \label{fig:isoc_all}
        \end{figure*}

        The results from the PARAM code are three-dimensional (3D) JPDFs of age, \omegai\, and \feh\ for each Cepheid. The JPDFs are returned by our Bayesian analysis, which is performed adopting 
        the value of $\sigma$ for each parameter given in Table~\ref{tab:stars_data}, for each star.
        
        Our first step is to check if the stars belong to the same population, hence we look at their metallicity and age marginalized 1D-PDFs. Figure~\ref{fig:combined_mpdf} shows the 1D-PDFs of age and metallicity of the stars in the left- and right-hand panels, respectively. The PDFs show a good agreement in both age and metallicity distributions; the only exception to this is the Cepheid HV12199. This star deviates from the trend of the others, and it seems to be older and more metal poor. Furthermore, we note that its 1D-PDF in age is clearly bimodal. As already anticipated (see Sec.~\ref{sec:Data_and_methods}), HV12199 is the less massive star in the sample by \citet{Marconi2013} and was found to be slightly more luminous than the luminosity predicted by the theoretical mass-luminosity relation (MLR) used for comparison in their work (see their Figure 10). These authors ascribed such over-luminosity to possible effects induced by mass loss.
        Assuming that all stars belong to the same population, i.e. same age and metallicity for all the stars, we may compute the combined PDFs (1D-CPDF) in age and metallicity with the product of the five 1D-PDFs.
        It is immediately apparent in the left-hand panel of Figure~\ref{fig:combined_mpdf} that the combined distribution in the case of the age is flat and almost zero at all ages. In other words, there are no models that can fit at once HV12199 and the other four Cepheids. On the contrary if we exclude HV12199, the combined 1D-CPDF shows a well-defined age peaking  at 176~Myr, as can be seen from the left panel of Figure~\ref{fig:combined_mpdf}. These four stars, which are younger than the ages provided by the 1D-PDFs of HV12199, are referred to in the following  as the young Cepheid population.
        As far as the metallicity is concerned, both the combined PDFs (i.e. the PDF computed with all the five stars and that computed excluding HV12199) peak at about $\feh=-$~0.35, indicating that the observed data are less sensitive to differences in the metal content. 
        This preliminary analysis already suggests that the Cepheid HV12199 may belong to a different population. 
        
        We now marginalize the 3D-JPDFs with respect to the metallicity to obtain the 2D-JPDFs on age and rotation. Then, we constrain the 2D-JPDFs of the young Cepheid population
        to follow their combined  four-star CPDF age distribution shown in the left panel of Figure~\ref{fig:combined_mpdf}.
        An example of the resulting 2D cJPDF of the Cepheid HV12198 is shown in the bottom-left panel of Figure~\ref{fig:HV19198_corr}. The coloured contours arbitrarily indicate chosen levels of the cJPDF, which are 50 per cent (the darker), 10 per cent (the intermediate), and 1 per cent (the lighter) of the maximum value of the cJPDF. The top-left and bottom-right panels show the 1D marginalized PDFs on \omegai\ and age, respectively. As suggested by \citet{Rodrigues2014}, we chose the peaks of the marginalized 1D PDFs (the mode, indicated by the dotted lines) as best values and we selected the smallest interval around the mode that contains the 68 per cent of the distributions as the credible intervals (CIs, indicated by the shaded areas). The best value and the CIs are represented by the square and the black error bars in the 2D JPDF plot (the bottom-left panel). The selected best value is not the peak of the 2D JPDF, however this point is inside the chosen CIs in each distribution.
        The only difference between the age PDF shown in the bottom-right panel and the combined four stars age distribution shown in Figure~\ref{fig:combined_mpdf} is that the first PDF has been obtained from the 2D cJPDF after the normalization to its maximum. Hence, while the absolute values are different, the shape, the derived best value, and the CIs of the two distributions are exactly the same.

        Table~\ref{tab:cep_results} lists the resulting best values and the corresponding CIs for the selected Cepheid, both for the non-corrected and the cJPDFs (except for the star HV12199, which will be further discussed below). The common age of the four Cepheids obtained is 176$_{-6}^{+4}$~Myr and the common metallicity is \feh~=~$-0.35_{-0.02}^{+0.01}$. We remark that these four young Cepheids have slow initial rotation rates (even taking into account the CIs). 
        Figure~\ref{fig:isoc_all} shows the comparison between the Cepheids data and selected isochrones in age, \omegai, metal content, and mass (values and CIs listed in Table \ref{tab:cep_results}). Each panel shows the best-fitting isochrone for the different stars, excluding HV12199, adopting the best value of \feh\ found in the analysis. The plot shows that the four Cepheids are all in the core helium burning (CHeB) phase. It is worth noting that the obtained small rotation rates are also in agreement with the ability of the corresponding models to produce more extended blue loops that can reach and cross the  instability strip, as discussed in Sec.~\ref{sec:Tracks&Isos} and illustrated in Figure~\ref{fig:tracks_cepheid}.
        \begin{figure}
            \includegraphics[width=0.45\textwidth]{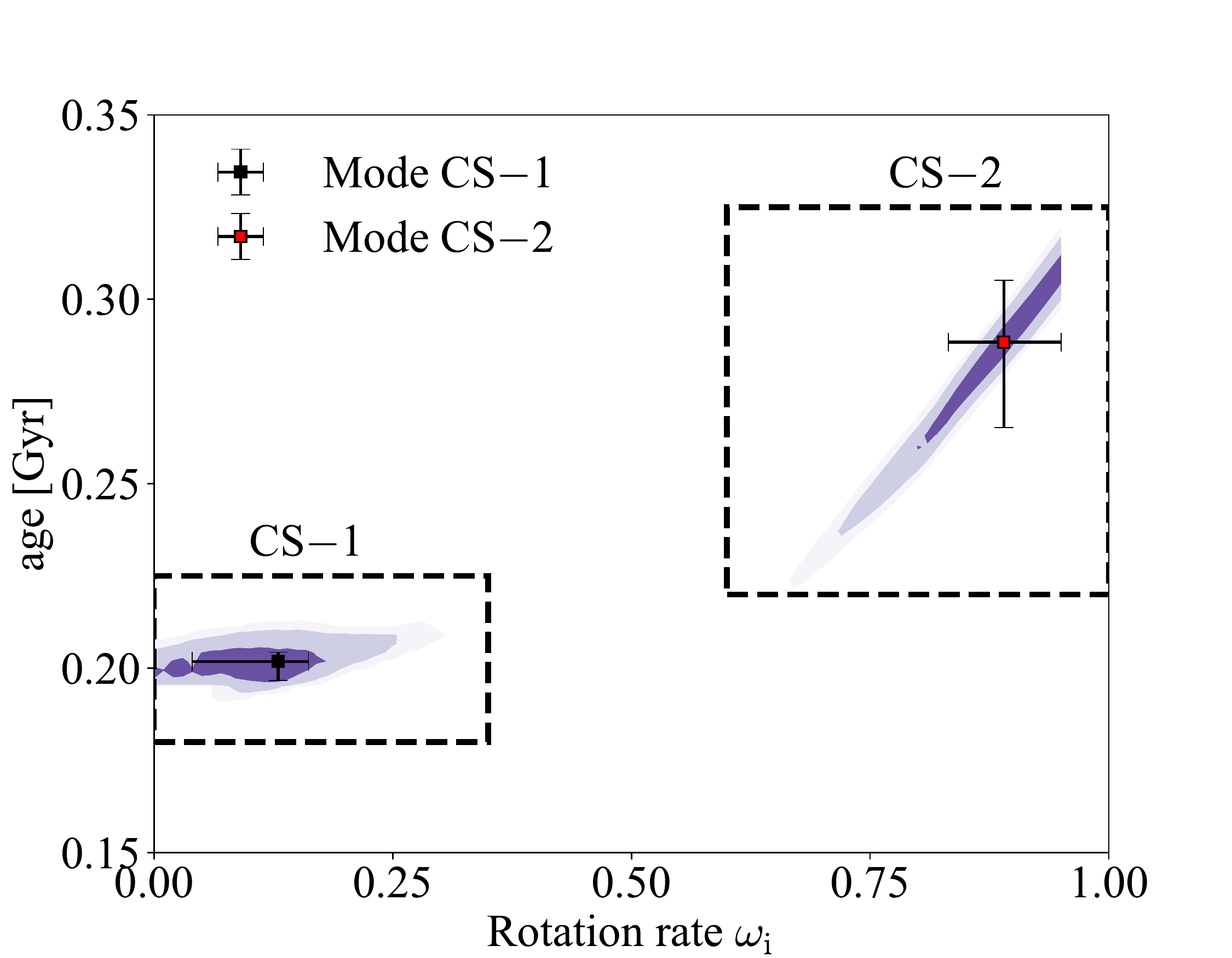}
            \caption{
                2D-JPDF as a function of age and initial rotation rate for the Cepheid HV12199. The contours are coloured for different levels of the JPDF as in Figure~\ref{fig:HV19198_corr}. The squares and error bars indicate the two mode values of the two classes of solutions.
                    }
            \label{fig:HV12199_singlemap}
        \end{figure}
        \begin{figure*}
            \includegraphics[width=0.45\textwidth]{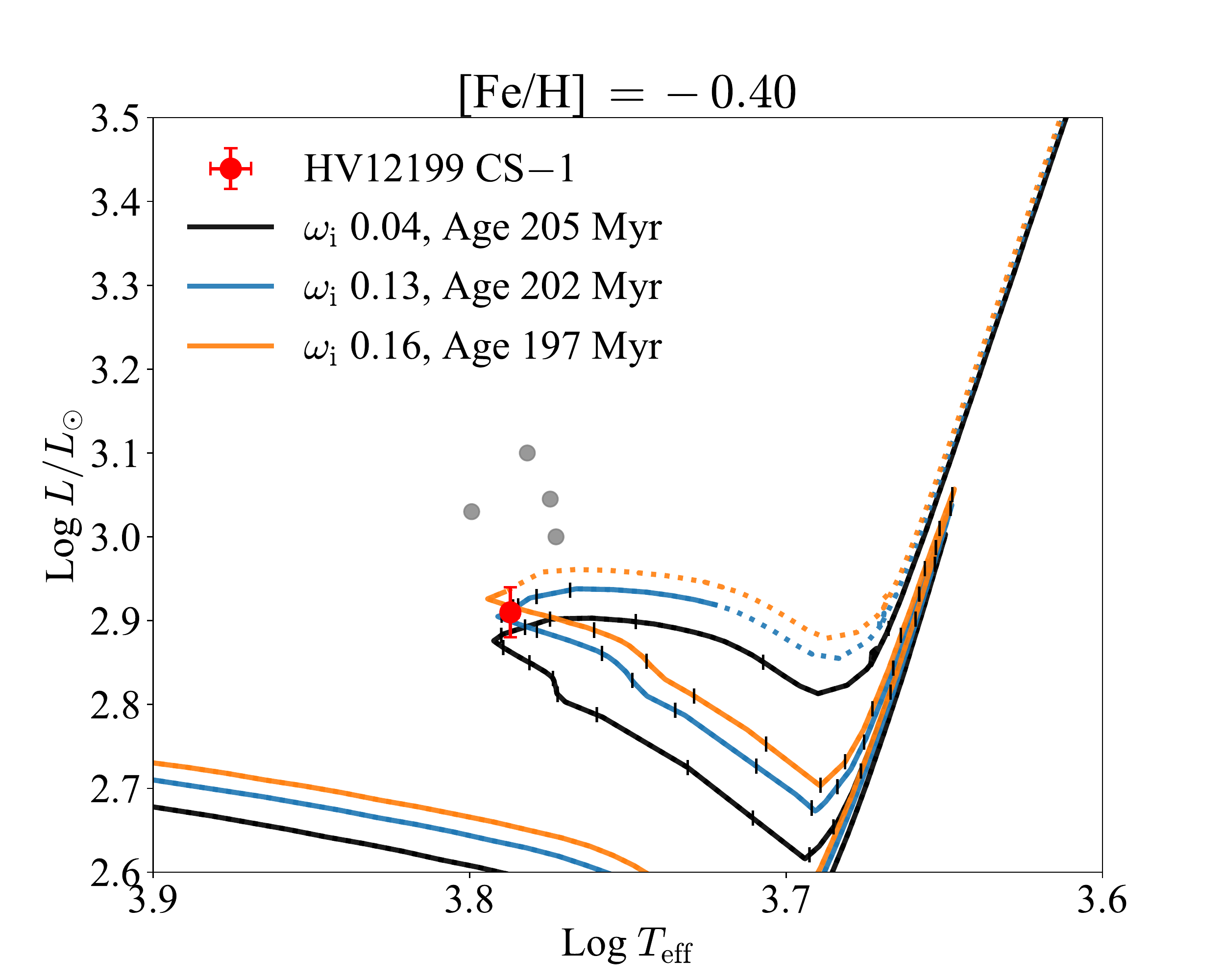}
            \includegraphics[width=0.45\textwidth]{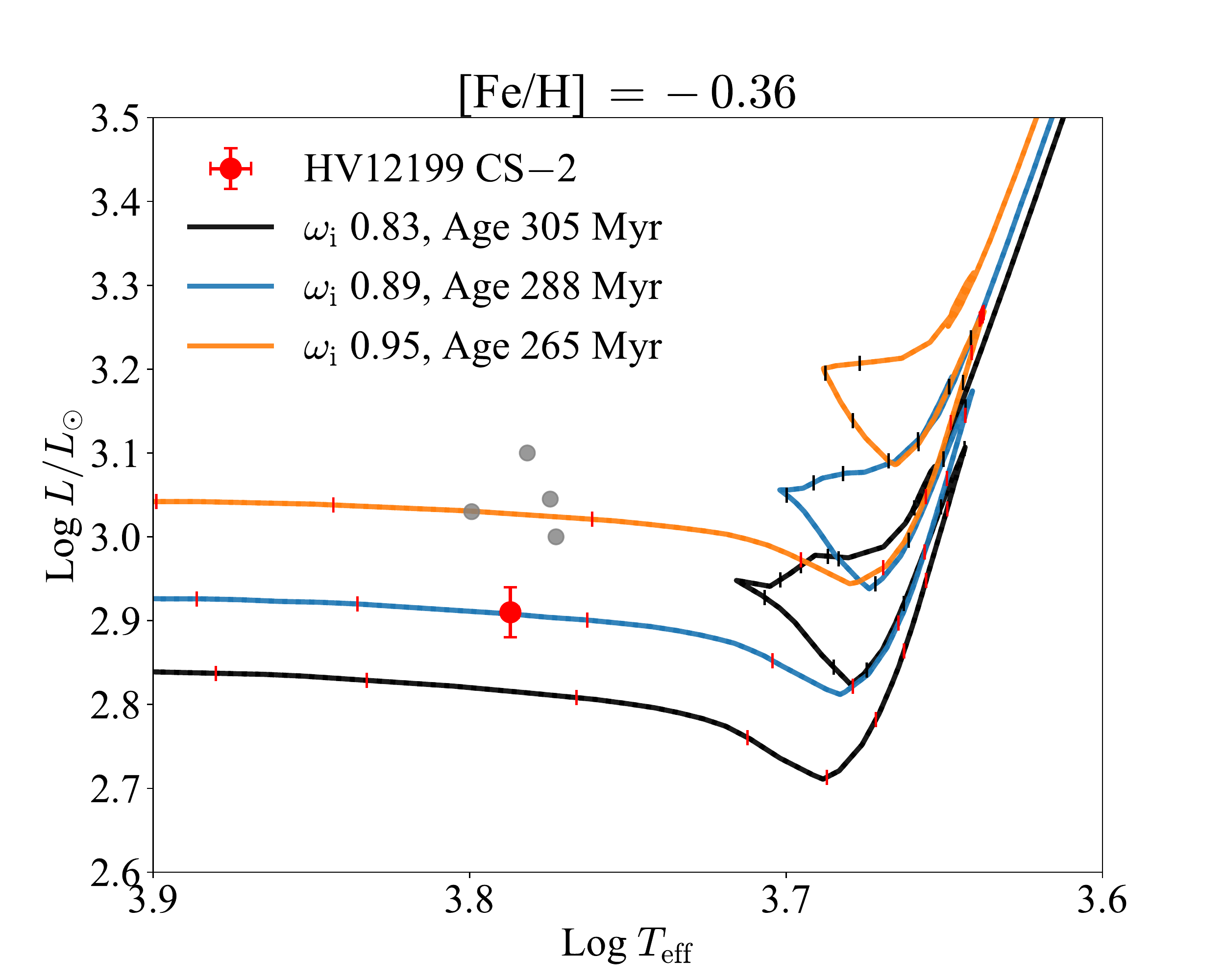}
            \caption{
                Comparison as shown in Figure~\ref{fig:isoc_all}. In the left-hand panel shows the first class of solution (CS-1) with isochrones with slow initial rotation rates and ages around 202~Myr. The right-hand panel shows the second class of solution (CS-2) with isochrones with high rotations and ages around 288 Myr. The values are selected from Table~\ref{tab:cep_results}. In the right-hand panel the red vertical markers in the Hertzsprung gap indicate intervals of $\Delta M$~=~0.0005~\Msun.
                    }
            \label{fig:isoc_HV12199}
        \end{figure*}
        \begin{figure}
            \includegraphics[width=0.45\textwidth]{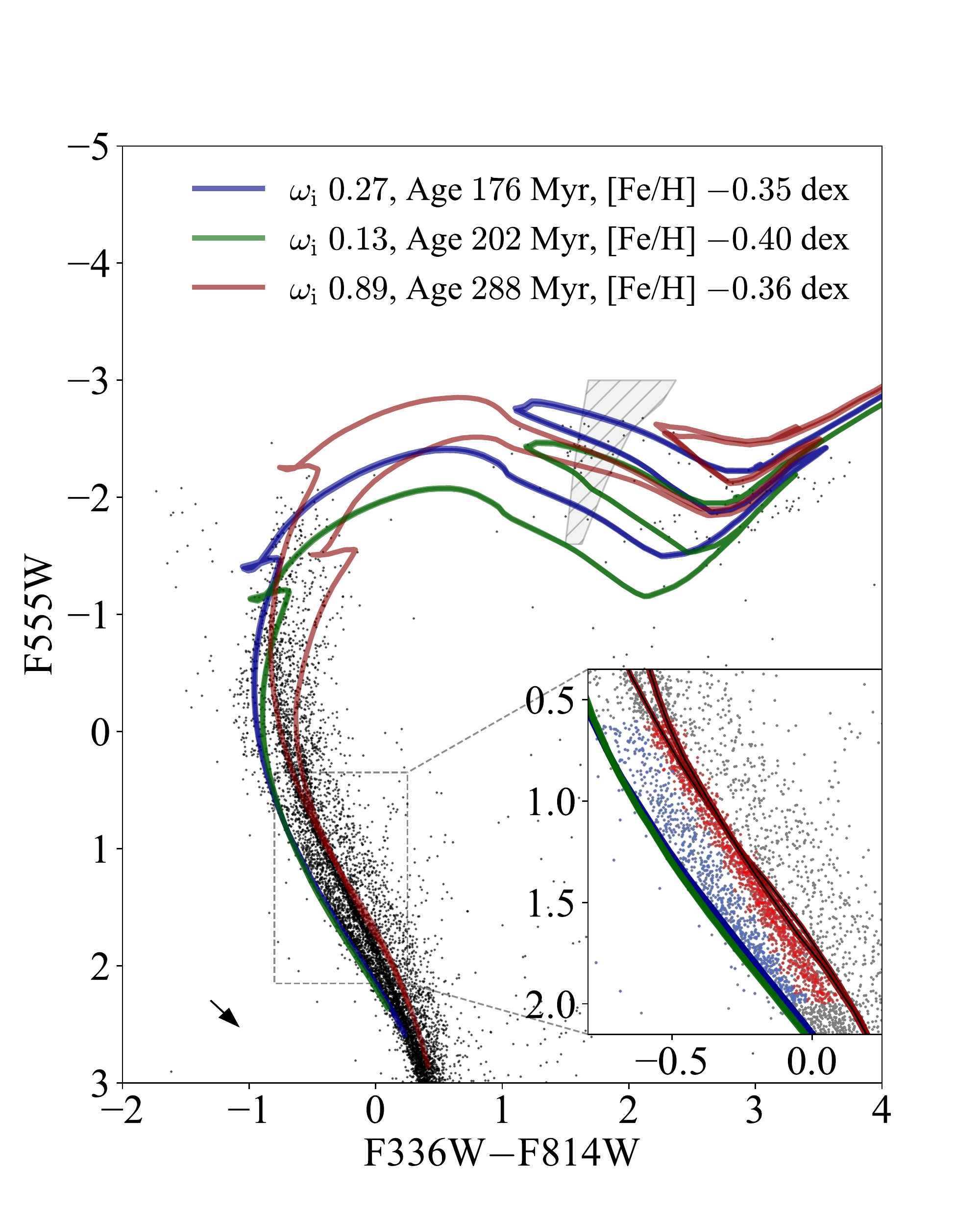}
            \caption{
                Comparison between the NGC~1866 cluster data and selected isochrones in the F555W vs. F336W~-~F814W CMD. The data description is the same as in Figure~\ref{fig:cmd}. The blue isochrone is selected taking the best values found from the common age and metallicity of the four-star combined PDF. The green and red isochrones are the two solutions of HV12199 with low (CS-1) and high initial (CS-2) rotation rates, respectively. All the isochrones are plotted with the effect of gravity darkening. The inclination angles, \textit{i}, of are 0$^\circ$ (pole on) and 90$^\circ$ (edge on) for the most and less bright isochrones, respectively. The shaded area indicates the Cepheids instability strip.
                    }
            \label{fig:cmd_fit1}
        \end{figure}

        As far as the star HV12199 is concerned, in Figure~\ref{fig:HV12199_singlemap} we show the 2D-JPDF, obviously not constrained by a common age. The 2D distribution represents two well-detached regions (or classes) of solutions, which are analysed separately. The first region (from now on CS-1) peaks at slow initial rotation rates with an age of about 200~Myr and metallicity of $-0.40$~dex (Z~$\sim$~0.006). The second region, instead, is centred at high \omegai\ and older ages ($\sim$290~Myr) and has a metallicity of \feh~=~$-$0.36. We refer to this solution as CS-2. The best values of the two classes and their corresponding CIs are shown in the bottom section of Table~\ref{tab:cep_results}. Both solutions give ages that are  different from the common age found for the young Cepheid population, in particular, the second case ($\sim$~290~Myr). As concerns the metallicities, CS-1 peaks at a lower \feh\ with respect to the common metallic content, while CS-2 is in agreement within the CIs.   
        
        In Figure~\ref{fig:isoc_HV12199} we compare the location of HV12199 in the HR diagram with the isochrones corresponding to the two different classes of solutions. The left-hand panel shows the isochrones selected for the CS-1 case, while the right-hand panel shows the isochrones selected for the CS-2 case. The isochrones are selected using the best value of \feh\ for each solution.
        In the CS-1 case HV12199 is in the CHeB phase, while the CS-2 case favours a position in the Hertzsprung gap phase. 
        This analysis explains why in Figure~\ref{fig:HV12199_singlemap}, and in the left-hand panel of Figure~\ref{fig:combined_mpdf}, the second solution (CS-2, with higher \omegai, cluster metallicity, and older ages) is less likely than the first (CS-1, with lower \omegai, lower metallicity, and younger ages). Indeed intermediate-mass stars spend the bulk of their post-MS lifetime in the CHeB phase. The other advanced evolutionary phases are much shorter. In particular a 4~\Msun\ star spends 
        only a few 10$^5$ years in the Hertzsprung gap phase, 
        which is about a factor ten shorter than the time spent on the CHeB phase.  
        We note however that in our Bayesian analysis we explore a parameter space of intervals of 3$\sigma$ for each value given in Table~\ref{sec:NGC1866_data} and, if we restrict the computation to only 1$\sigma$ around the observed values, the CS-1 solution disappears leaving only the CS-2 solution.
That said, we have kept both solutions and now discuss the fit of the CMD data.

    \subsection{Colour-magnitude diagram}
    \label{sec:res_cmd}

        In this Section we analyse the observed CMD of NGC~1866 in the light of our previous results. In particular we compare isochrones with ages, metallicities, and initial rotation rates derived from the analysis of the Cepheids properties with the CMD of the NGC 1866 cluster. We stress that our comparison is not the result of a best-fit procedure to the observed data. Instead, we simply superimposed the selected isochrones to the data adopting a distance modulus $(m - M)_0 = 18.43$ mag and an extinction $A_V = 0.28$ mag derived by a previous independent fit performed with PARSEC V1.2s \citep{Goudfrooij2018}.
        We note that the distance modules found by \citet{Goudfrooij2018} using the superior \HST\ photometry and the new evolutionary tracks are slightly smaller than those derived by \citet{Marconi2013}, who found $(m - M)_0 = 18.56 \pm 0.03$(statistical) $\pm \, 0.1$(systematic) mag. But these results are still within their quoted total error. We return to this point in the discussion section. 

        We used a young population of slowly rotating stars with age of 176~Myr and metallicity \feh~=~$-$0.35, corresponding to the young Cepheid population. The initial rotational velocity is the mean value determined for the four Cepheids, \omegai~=~0.27  (see Table~\ref{tab:cep_results}). 
        As for the Cepheid HV12199, we used isochrones corresponding to the  CS-1 and CS-2 solutions. For the CS-1 solution we adopted \omegai~=~0.13, age of 202~Myr and metallicity \feh~=~$-$0.40. The CS-2 solution is represented by \omegai~=~0.89, age of 288~Myr and metallicity \feh~=~$-$0.36. 
        In all cases the isochrone are  plotted for two extreme values of the inclination angles, $i = 0^{\circ},\,i = 90^{\circ}$. However, as shown in Figure~\ref{fig:isoc_gravity}, the effects of the gravity-darkening below a rotation rate of 0.6 are almost negligible, hence even though these effects are taken into account and plotted in Figure~\ref{fig:cmd_fit1}, they  produce indistinguishable results in the slowly rotating isochrones. 

        Figure~\ref{fig:cmd_fit1} shows the comparison of the
        isochrones described above with the observed data in the CMD of NGC~1866.
        As can be seen from the figure, the isochrones representing the young Cepheid population (176~Myr, \feh~=~$-$0.35, $i = 0^{\circ},\,i = 90^{\circ}$) nicely reproduces the bluest part of the MS. We also see that the bluest part of the MS may be well reproduced by the isochrone that corresponds to the CS-1 solution. 
        We note that, as expected, the CS-1 isochrone does not reproduce the location of the other four Cepheids simultaneously. 
        The isochrone selected from the CS-2 solution reproduces very well the turn-off of the cluster and the red MS but, as it shown from the inset, the lower red MS is not reproduced perfectly. 
        
        In summary, we have shown in this Section that using stellar populations with parameters determined by the Bayesian analysis of well-studied Cepheids in NGC 1866, we can nicely reproduce the main features of its CMD. 
        Better fits could likely be achieved using isochrone parameters that represent a range of \omegai\ and/or \feh\ values selected within the entire CIs of the solution.
        This will be carried out in a future investigation together with full CMD simulations. 

%
%

\section{Discussion and conclusions}
\label{sec:Disc_and_conc}

    In this paper, we analyse the evolutionary properties of five Cepheids in the young cluster NGC~1866 with extensively studied pulsational properties. A Bayesian analysis based on new grids of evolutionary tracks shows that it is very unlikely that all the five Cepheids belong to the same stellar population. 
    The age distribution of HV12199 obtained with normal single star evolutionary models is clearly differentiated from that of the other four Cepheids. In order to bring HV12199 in agreement with the other four Cepheids it should have lost about 0.5-0.7~\Msun. The mass lost is too large in the case of single stellar evolution, as shown by the other four stars. In the case of binary evolution, since the the external envelope of a 4~\Msun\ is of about 3~\Msun, the lost mass would require a fine tuning of the binary parameters. These considerations led us to the conclusion that the two group of Cepheids are representative of two different populations harboured by NGC~1866.  We find that four Cepheids are likely descendants of a population of initially slow rotating stars (with \omegai~$\leq$~0.4), which have an age of 176$^{+4}_{-6}$~Myr and a \feh~=~$-$~0.35$^{+0.01}_{-0.02}$. 
    Instead, for HV12199 we find that possible solutions are either a 202$^{+3}_{-5}$~Myr old slowly rotating population with \feh~=~$-$~0.40$^{+0.01}_{-0.03}$, or a 288$^{+17}_{-23}$~Myr old and initially fast rotating (\omegai~$\sim$~0.90) population with \feh~=~$-$~0.36$^{+0.02}_{-0.04}$, named CS-1 and CS-2, respectively. 
    
    The existence of such a range of ages for NGC~1866 Cepheids is not new.
    In fact, while  \citet{Musella2016} favoured a single population with an average age of 140~Myr, significant age differences have been already reported in literature. \citet{Lemasle2017}, using the period-age relations of non-rotating \citet{Bono2005} models, found ages between 95~Myr and 115~Myr, while \citet{Anderson2016} found ages between 180~Myr to 250~Myr using rotating models. These differences may be due to the various prescriptions adopted for the non-rotating models (in their case without the core overshooting), and for the rotating models (they adopted a velocity of \omegai~=~0.5). Interestingly we note that the age found in their work for HV12198 (about 184~Myr) adopting \omegai~=~0.5, is very close the common age (176$^{+4}_{-6}$~Myr) we adopted in this paper for our four slowly rotating Cepheids. 

    Remarkably, the observed CMD has clear signatures of multiple stellar populations. The presence of at least two different stellar populations in NGC~1866 with different rotational properties has been already suggested, for example, by \citet{Milone2017, Goudfrooij2018, Girardi2019}, and  corroborated even by direct spectroscopic observations of the stars in the eMSTO \citep{Dupree2017}.
    In this work, we find that the isochrone corresponding to the slowly rotating young Cepheids population reproduces the blue\ MS of NGC~1866 almost perfectly.
    The isochrone corresponding to the fast rotating population derived for  HV12199 (CS-2) reproduces very well the observed red MS in the CMD. Near the cluster turn-off this sequence widens into a strip that is nicely fitted by taking into account the gravity darkening effects caused entirely by the relative inclination of the rotation axis with respect to the line of sight. The alternative case of a slow initial rotation for HV12199 (CS-1) corresponds to an isochrone which is almost superimposed to the blue sequence. We stress once again that the isochrones shown in Figure~\ref{fig:cmd_fit1} are not best fits to observed sequences but simple plots of the best solutions found for the Cepheids.

    A striking feature in the CMD of NGC~1866 is that the red MS becomes bluer than the model isochrone at decreasing luminosity. Since these stars are not significantly evolved the differences between non-rotating and fast rotating models must originate from the geometrical effects alone. This could be an indication that these effects are not yet properly modelled or, on the contrary, that the initial rotational properties depend on the stellar mass \citep{Goudfrooij2018}.
    
    Concerning the inclination angles of fast rotators in NGC~1866, we note that in a recent analysis of the two open clusters NGC~6791 and NGC~6819, \citet{Corsaro2017} found a strong alignment between the cluster rotation axes and the stars rotation axes, thus claiming that this could have been a general feature in star clusters. However this was not confirmed by more recent asteroseismic studies of the same clusters \citep[performed by][]{Mosser2018} where, instead, a stochastic dispersion of the stellar inclination angles has been found. We have seen in this work that the eMSTO of NGC~1866  is very well explained by different inclinations of the old fast rotating stars, which agrees with findings by \citet{Mosser2018}.
    
    The Bayesian analysis shows that, at maximum, only one of five Cepheid studied descends from  initially fast rotating stars. It is interesting to see whether this is in contrast with  the claim that about two-thirds of the MS stars in NGC~1866 are fast rotators with \omegai~=~0.90 \citep{Milone2017}. To this purpose we need to properly account for the  evolutionary time spent by the stars within the instability strip. 
    In fact, at turn-off masses typical of NGC~1866,  evolutionary tracks with high initial rotations have much less extended blue loops and  are not able to cross the Cepheids instability strip during central He burning
    (see Figure~\ref{fig:tracks_cepheid}). In this case  the tracks cross the instability strip only during the Hertzsprung gap in  timescales that are much shorter than those of the CHeB phases, disfavouring the  Cepheids phase. To better clarify this point we quickly estimated the relative number of Cepheids expected from two stellar populations corresponding to thoses highlighted by our analysis. 
    We used the two isochrones representative of the slow rotating young population (with  \omegai~=~0.27 and an age of 176~Myr) and of the the fast rotating old population (the CS-2 solution, with \omegai~=~0.89 and 288~Myr). 
    The number of stars in a the Cepheid phase can be expressed as 
    \begin{equation}
         N \sim \sum^{K}_{j=1} \int_{\Delta M_J} \Phi(\Mi) \; \delta \Mi
    ,\end{equation}
    where $K$ is the number of times that the isochrone crosses the instability strip, $\Delta M_J$ is the interval of mass within the strip for each cross, $\Phi(\Mi)$ is the IMF, and \Mi\ is the initial stellar mass.
    Using a \citet{Salpeter1955} IMF and taking into account that, following \citet{Milone2017}, the ratio of the two populations in the MS is one-half, we find that
    the ratio of the fast initially rotating to the slow rotating Cepheids expected in NGC~1866 is  $N_\mathrm{fast}/N_\mathrm{slow} \sim 1/20$; we hence show that initially non-rotating Cepheids are by far more likely to be observed.
    Thus the paucity of initially fast rotating Cepheids with respect to the non-rotating Cepheids (1/5) we found, is not in contrast with the  results of \citet{Milone2017}, once proper accounting is done of the evolutionary time spent within the instability strip. 

    We already anticipated that the distance modulus adopted in this work, obtained by a fitting procedure on more recent \HST\ photometry of the cluster, is about 0.13~mag smaller than that obtained by \citet{Marconi2013}. However, both distance moduli agree within their total errors, i.e. within the sum of the systematic and the statistic errors provided.
    We note that the errors adopted in our Bayesian analysis correspond to three times the statistic errors provided by \citet{Marconi2013} for the structural parameters of the Cepheids and, from our analysis, 
    we obtain ages that consistently fit the MS split and eMSTO by adopting the shorter distance modulus (see Fig.~\ref{fig:cmd_fit1}).
    It is also worth recalling that the mass difference between the Cepheid HV12199 and the others is independent of the distance modulus obtained by \citet{Marconi2013}. 
    
    We stress that to be fully consistent, the pulsational analysis should be performed using the same input physics adopted in our evolutionary models, which however is beyond the goal of this paper.
    To get a first order estimation of the possible effects due to the adoption of different models for the pulsational analysis, we make use of the period-mass-luminosity-effective temperature ($PML$\Teff) relation by \citet[][ their equation 4]{Chiosi1993}. By differentiating this relation and assuming that the uncertainties in the period and the \Teff\ are negligible \citep[see Tables 1 and 2 of][]{Marconi2013}, we obtain that $\frac{\delta \log M}{\delta \log L}\sim 1$. For example, a variation of the luminosity of say, 10 per cent, owing to different input physics, such as a different adopted mixing scheme and different opacities, should correspond to a mass estimation that differs of the same order.

    As a final consideration we note that from a Bayesian analysis of Cepheids data based on a large grid of stellar models with varying initial masses, rotational velocities, and  metallicities, we are able to obtain the ages of the two main populations of NGC~1866  that clearly stand out in its CMD. 
    The older population is composed of fast rotating stars, while the younger population is comprised of slowly rotating stars. Their metallicity is almost identical.
    First of all this finding shows that, in agreement with \citet{Milone2017} and \citet{Dupree2017}, the observed properties of NGC~1866, and likely of similar clusters, result from a complex mixture of physical effects that include both rotation and age dispersion. A combination of these two effects  seems to be a necessary ingredient to correctly interpret the clusters formation and evolution. 
    Second, the characteristics of the main stellar populations suggest a well-defined evolutionary scenario for the cluster. A first burst generated the older population, which inherited the initial angular momentum from the progenitor clouds, and now hosts the biggest fraction of the initial global angular momentum content. After about 130~Myr another generation of stars forms out of the gas that already has the lost memory of the initial angular momentum. The stars of this younger generation are thus mainly slow rotators. The metallicity remains almost unchanged   as can be derived from the \feh\ content of the two populations.  A similar scenario has been already suggested by \citet{Dupree2017}, however with a significant difference in the age of the older population that in their case is of about 200~Myr instead of our finding of 290~Myr. 
    It is also interesting to note that this scenario has similarities with that suggested for the formation of multiple populations in old globular clusters by \citet{decressin2007} and \citet{charbonnel2013}.  In short, these authors suggested that the anomalies presently observed in the low mass stars of old GCs result from the ejecta, enriched in H-burning products, of a first generation of fast rotating massive stars. Indeed, a similar population of fast rotating massive stars could also have been present in NGC~1866, but what remains at the present age is just its related population of intermediate-mass stars, which are clearly older than the non-rotating population. Such a first generation of fast rotating stars, if confirmed in more clusters, could provide an interesting avenue for the interpretation of multiple populations as a whole, including the old globular clusters.
    
    To improve our analysis, a first step would be to enlarge the sample of well-studied Cepheids in NGC~1866, possibly exploiting the  full population of about 24 objects. 
    Furthermore, given the age differences found by different authors for the stellar populations in NGC~1866, it is also important to compare the different models and check whether these differences can be ascribed only to models or to some different initial assumptions as well, such as the amount of convective overshooting,  efficiency of the rotational mixing, or possible presence of the magnetic braking.
    
    At the same time it could be worthy to perform a full analysis of the CMD of this and other similar clusters in an attempt to better constrain the distributions of initial rotational velocities, star spin alignments, ages, and metallicities in these clusters.
All those considerations call for further efforts from both  the observational and theoretical sides. 

\begin{acknowledgements}
    We thank M.A.T. Groenewengen for the helpful comments. We acknowledge the support from the  ERC Consolidator Grant funding scheme ({\em project STARKEY}, G.A. n. 615604). AB acknowledges support from PRIN-MIUR~2017. TSR acknowledges financial support from Premiale 2015 MITiC (PI B. Garilli). For the plots we used matplotlib, a Python library for publication quality graphics \citep{Hunter2007}.
\end{acknowledgements}

 
\bibliographystyle{aa} 
\bibliography{ms} 

\end{document}